\documentclass[journal]{IEEEtran}

\usepackage{cite}
\usepackage{array}
\usepackage{color}
\usepackage{float}
\usepackage[cmex10]{amsmath}
\usepackage{nomencl}						
\usepackage[normalem]{ulem}			        
\usepackage{fixltx2e} 					    
\usepackage{diagbox}						
\usepackage{slashbox}						
\usepackage{colortbl}						
\usepackage{multirow}						
\usepackage{tabularx}
\usepackage{placeins}
\usepackage{algorithm}
\usepackage{mathrsfs}
\usepackage{mathdots}
\usepackage{amssymb}
\usepackage{arydshln}
\usepackage{color}
\usepackage{gensymb}
\usepackage[noend]{algpseudocode}
\usepackage{bm}
\usepackage{url}
\usepackage[hidelinks]{hyperref}
\usepackage[T1]{fontenc}
\newcommand{\subparagraph}{}
\usepackage{fancyhdr} 

\ifCLASSINFOpdf
  \usepackage[pdftex]{graphicx}
	\graphicspath{{./figure/}}
  \DeclareGraphicsExtensions{.pdf,.jpeg,.png}
\else
  \usepackage[dvips]{graphicx}
  \graphicspath{{./figure/}}
  \DeclareGraphicsExtensions{.eps}
\fi

\ifCLASSOPTIONcompsoc
  \usepackage[caption=false,font=normalsize,labelfont=sf,textfont=sf]{subfig}
\else
  \usepackage[caption=false,font=footnotesize]{subfig}
\fi

\newcommand{\figref}[1]{\figurename~\ref{#1}}

\ifCLASSOPTIONonecolumn

\else

\fi

\begin{document}
\bstctlcite{IEEEexample:BSTcontrol}
\setcounter{page}{1}

\title{Impedance-Based Whole-System Modeling for a Composite Grid via Embedding of Frame Dynamics}

\author{Yunjie~Gu, \IEEEmembership{Senior Member, IEEE}, Yitong~Li, \IEEEmembership{Student Member, IEEE}, Yue Zhu, Timothy~C.~Green, \IEEEmembership{Fellow, IEEE}
\thanks{Yunjie Gu, Yitong Li, Yue Zhu, and Timothy C Green are with the Department of Electrical and Electronic Engineering, Imperial College, London. E-mail: \url{yunjie.gu@imperial.ac.uk}; \url{yitong.li15@imperial.ac.uk}; \url{yue.zhu18@imperial.ac.uk}; \url{t.green@imperial.ac.uk}.}
\thanks{This work was supported by the Engineering and Physical Sciences Research Council of UK (EPSRC) under awards EP/S000909/1.}}

\ifCLASSOPTIONpeerreview
	\maketitle 
\else
	\maketitle
\fi

\thispagestyle{fancy}
\lhead{IEEE TRANSACTIONS ON POWER SYSTEMS}
\rhead{\thepage}
\cfoot{}
\renewcommand{\headrulewidth}{0pt}
\pagestyle{fancy}
\lhead{IEEE TRANSACTIONS ON POWER SYSTEMS}

\begin{abstract}
The paper establishes a methodology to overcome the difficulty of dynamic frame alignment and system separation in impedance modeling of ac grids, and thereby enables impedance-based whole-system modeling of generator-converter composite power systems. The methodology is based on a frame-dynamics-embedding transformation via an intermediary steady frame between local and global frames, which yields a locally defined impedance model for each generator or converter that does not rely on a global frame but retains all frame dynamics. The individual impedance model can then be readily combined into a whole-system model even for meshed networks via the proposed closed-loop formulation without network separation. Compared to start-of-the-art impedance-based models, the proposed method retains both frame dynamics and scalability, and is generally applicable to various network topologies (meshed, radial, etc) and combinations of machines (generators, motors, converters, etc). The methodology is used to analyze the dynamic interaction between generators and converters in a composite grid, which yields important findings and potential solutions for unstable oscillation caused by PLL-swing coupling in low-inertia grids.
\end{abstract}


\setlength{\nomitemsep}{0.4em}
\makenomenclature
\nomenclature[01]{$dq$}{Local swing frame}
\nomenclature[02]{$d^\prime q^\prime$}{Local steady frame}
\nomenclature[03]{$\alpha\beta$}{Stationary frame}
\nomenclature[04]{$\delta$}{Angle difference between a local swing frame and the global swing frame}
\nomenclature[05]{$\xi$}{Angle difference between a local steady frame and the global steady frame}
\nomenclature[06]{$\epsilon$}{Angle perturbation between a local swing frame and a local steady frame}
\nomenclature[07]{$\theta$}{Angle displacement}
\nomenclature[08]{$\omega$}{Angular speed (frequency)}
\nomenclature[09]{$s$}{Laplace operator}
\nomenclature[10]{$i$}{Current}
\nomenclature[11]{$v$}{Voltage}
\nomenclature[12]{$\psi$}{Flux-linkage}
\nomenclature[13]{$L$}{Inductance}
\nomenclature[14]{$R$}{Resistance}
\nomenclature[15]{$C$}{Capacitance}
\nomenclature[16]{$Z$}{Impedance}
\nomenclature[17]{$Y$}{Admittance}
\nomenclature[18]{$K_i$}{Current governed frame dynamics}
\nomenclature[19]{$K_v$}{Voltage governed frame dynamics}
\nomenclature[20]{$T$}{Torque}
\nomenclature[21]{$D$}{Damping torque coefficient}
\nomenclature[22]{$J$}{Rotor inertia}

\ifCLASSOPTIONonecolumn
	\printnomenclature[2cm]
\else
	\printnomenclature
\fi

\section{Introduction}

The major power systems of the world are transforming from ones dominated by sources employing synchronous generators (e.g. hydro, natural gas, and nuclear) to composite systems in which synchronous generators and power-electronic converters interfaced sources (e.g. wind, solar, battery) co-exist. Such a composite grid gives rise to new dynamic behaviors and new threats to system stability, and calls for new approaches to system modeling and analysis to address the problems. 

The most generic approach to model a composite grid is the state-space method \cite{kundur1994power,2016WDu,2016YWang}, which covers all internal dynamics of every machines (including generators, converters, transmission and distribution facilities, and loads) in the grid, and offers insights into the root cause of each oscillation mode via participation analysis \cite{4494608,260827,7526438,gu2018reduced}. However, state-space models rely on detailed knowledge of hardware design and control algorithms of each machine, which may not be available for power electronic converters since they are not usually disclosed by manufacturers nor standardized across different suppliers.  

Impedance (or equivalently, admittance) models are considered to be a useful alternative to the state-space method \cite{harnefors2007input, wang2014modeling, wang2018harmonic}. An impedance is essentially a transfer function description (from current input to voltage output, or vice versa for admittance) of an electrical system. Such a transfer function does not represent internal details or states, and can be measured or validated where high-fidelity analytical models are not available. Impedance models prove to preserve all information concerning dynamic behavior (in the sense of small-signal analysis) provided that every state is controllable and observable from the port at which the impedance model is formed and therefore such an impedance model can produce results for system stability analysis equivalent to those of the state-space method \cite{amin2017small}.

The impedance-based method of stability analysis has had widespread application in low-voltage dc power systems \cite{liu2003stability,sudhoff2000admittance,middlebrook1976input}. However, two fundamental difficulties arise when applying it to large-scale ac power systems. The first difficulty is dynamic frame alignment. Machines (converters and generators) in ac systems are usually modeled in a rotating reference frame aligned to a local angle such as the angle of the physical rotor or the angle identified by a phase-locked loop (PLL). These locally referenced models then need to be aligned, i.e. rotationally transformed, to a global reference frame to model the whole system. This frame alignment is straightforward in state-space models since all angles are explicitly represented in the states and therefore available to be used in rotational transformations between the local and global frames \cite{pogaku2007modeling}. The impedance model, on the other hand, does not explicitly include a representation of angle information since the input and output only contain voltage and current signals. As a result, impedance-based methods in ac systems have been limited to harmonic analysis in which all machines in the system are assumed to be synchronized to a fixed fundamental frequency with preset phase angles \cite{2019XCai,xu1991synchronous}. This approach is appropriate for conventional power systems with high inertia since the frame dynamics (i.e. rotor swing and frequency perturbation) in such systems is well separated in time-scale from harmonic terms. However, the boundary between frame dynamics and harmonics is becoming blurred recently because of reducing inertia in the system and the emergence of new synchronizing mechanisms (PLLs). Because of this blurring, a method able to model them together in a unified framework is now needed. A possible way to meet this requirement is to model every machine in a global reference frame where the frame dynamics appear as explicit paths in the impedance diagram \cite{2016Sync}. However, this procedure is rather complicated and not scalable, and therefore is limited to simple systems such as a single-converter-infinite-bus system or a stand-alone single-converter-load system \cite{2017BWen,cespedes2014impedance,sun2011impedance,gu2019motion}. 

The second difficulty is system separation. When using impedance-based models for system stability analysis, the system needs to be separated into source-side and load-side, so that the closed-loop transfer function can be formulated according to the impedance ratio between the source and load. Such separation is straightforward in low-voltage dc power system due to its radial and single-source topology. A large-scale ac power system, on the other hand, usually has a meshed and multi-source topology. Although it is possible to separate a meshed network according to its critical cut-set, finding a critical cut-set is more difficult than using the impedance model itself \cite{2019XCai,padiyar2013structure}. 

In this paper, a methodology is established to overcome these two difficulties which are blocking the use of impedance-based modeling and analysis in large-scale composite ac grids. To solve the dynamic frame alignment problem, an impedance transformation is defined to embed frame dynamics into local impedance models so that they can be aligned to the global frame without loss of any dynamic features. Compared to impedance modeling directly in the global frame \cite{2016Sync}, the proposed methodology retains scalability, that is, the impedance of each machine can be modeled or measured locally without referring to the external global frame, and the local impedances can be interconnected in a modular fashion to form a whole-system model. To solve the system separation problem, the system is reformulated as the closed-loop connection of machine impedance matrix (to represent both the sources and loads) and nodal admittance matrix (to represent the network). Such a formulation avoids the need of system separation and offers more flexible selections of the points where impedances or admittances are formed. The proposed methodology is generally applicable to a composite ac grid with different typologies (radial and meshed) and miscellaneous machines (synchronous generators, induction generators/motors, and power electronics converters). This methodology also features a scalable and standardized procedure which can be automated in a computer program and easily used by practitioners. 

The paper is organized as follows. The methodology of frame alignment in impedance-based whole-system modeling is introduced in Section II. The impedance transformation to embed frame dynamics into local impedance models is defined in Section III. The closed-loop formulation without system separation is discussed in Section IV. In Section V, the proposed methodology is used to investigate generator-converter interaction in a composite low-inertia grid to demonstrate the advantage of impedance-based whole-system analysis. Simulation results are given in Section VI. The last section concludes the paper.

\section{Frame Alignment in Whole-System Modeling}

\begin{figure}
\centering
\includegraphics[scale=1.0]{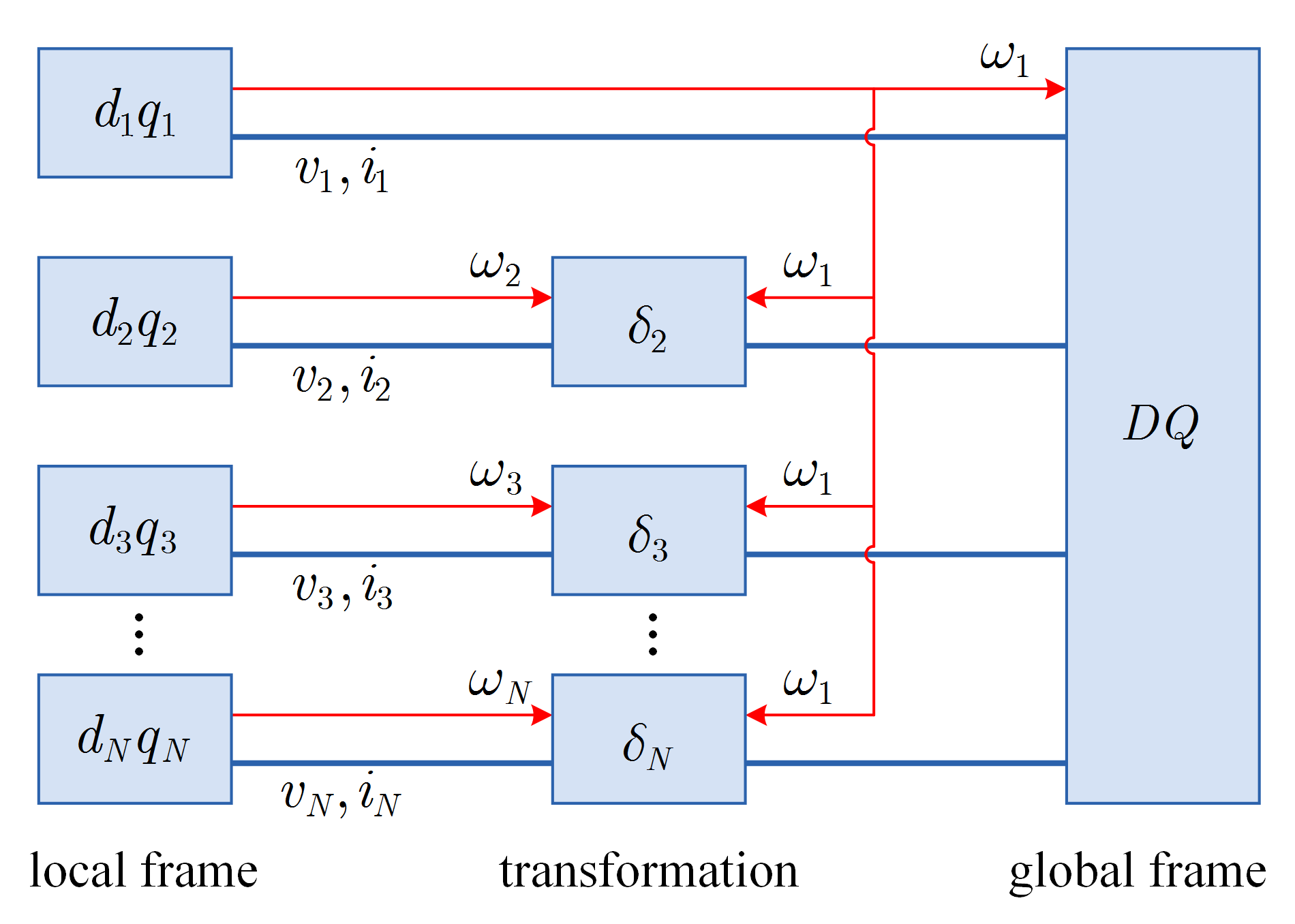}
\caption{Frame alignment in state-space modeling: frame frequency $\omega_n$ and angle $\delta_n$ are expressed by the states and readily available for frame transformations.}
\label{fig_align_ss}
\end{figure}

We start by reviewing how frame alignment is performed in state-space models and from this demonstrate why frame alignment is difficult in impedance-based models. For the state-space case in \figref{fig_align_ss}, each machine in the system is modeled in its local frame $d_n q_n$ $(n = 1,2,\cdots,N)$ first. The angle difference $\delta_n$ between a local frame and the global frame is determined by the state equation below
\begin{equation}
\label{eq_delta} 
\dot{\delta}_n = \omega_n - \omega_1
\end{equation}
in which $\omega_n$ is the local frame speed of $d_n q_n$ governed by each rotor or PLL, and $d_1 q_1$ is selected as the global frame. Since $\omega_n$, and therefore also $\delta_n$ are expressed in the states of the machines, the local-to-global frame transformation in a state-space model is straightforward. In an impedance model, however, only current and voltage are expressed, so there is no straightforward ways to identify the angles for local-to-global frame transformations. 

\begin{figure}
\centering
\includegraphics[scale=1.0]{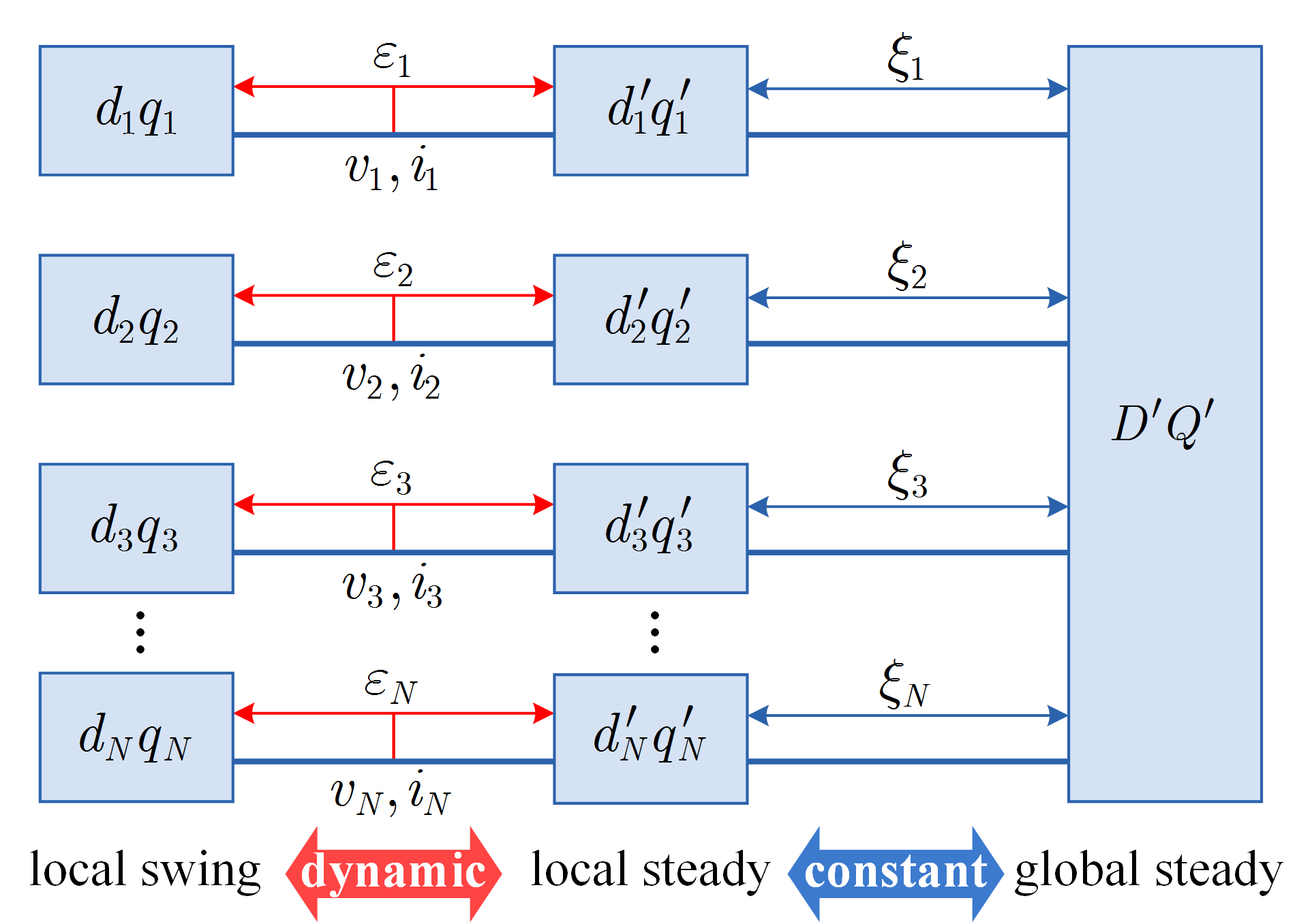}
\caption{Frame alignment in impedance modeling: frame dynamics introduced via dynamic angle $\epsilon_n$.}
\label{fig_align_imp}
\end{figure}

\begin{figure}
\centering
\includegraphics[scale=0.9]{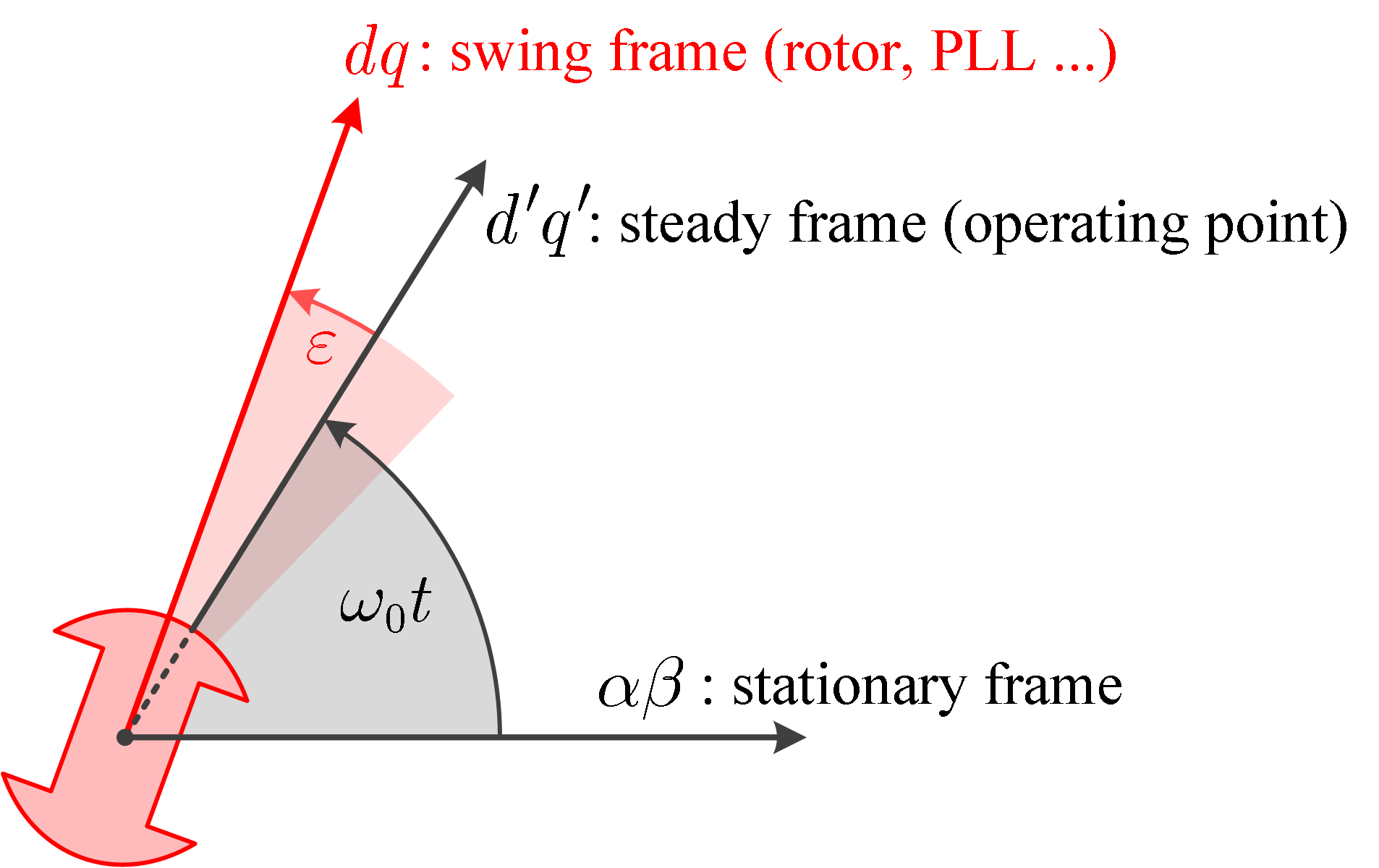}
\caption{Illustration of a swing frame $dq$ (such as the rotor frame of a synchronous generator) which swings around a steady frame $d^\prime q^\prime$ by a dynamic angle $\epsilon$.}
\label{fig_angle}
\end{figure}

To solve this problem, a method is proposed to embed frame dynamics (that is, the dynamics of $\delta_n$) into impedances. Firstly, the local frame is split into two sub-frames, namely the steady frame and swing frame. The steady sub-frame is defined as the steady-state operating point of the local frame: it rotates at a constant speed $\omega_0$ and has no dynamics. The swing sub-frame, on the other hand, is aligned to the instantaneous rotor or PLL angle, and can vary (or swing) around the steady sub-frame by an angle $\epsilon$, as illustrated in \figref{fig_angle}. 

The local steady sub-frame serves as an intermediary frame between a local frame and the global frame. Based on this intermediary frame, the impedance transformation from local to global can be performed in two steps: (i) from a local swing frame to a local steady frame, and (ii) from a local steady frame to the global steady frame, as shown in \figref{fig_align_imp}. This two-step transformation separates the global (constant) and local (dynamic) part of the frame alignment, and thereby retains both frame dynamics and scalability in the impedance model.

\begin{figure}
\centering
\subfloat[]{\includegraphics[scale = 0.95]{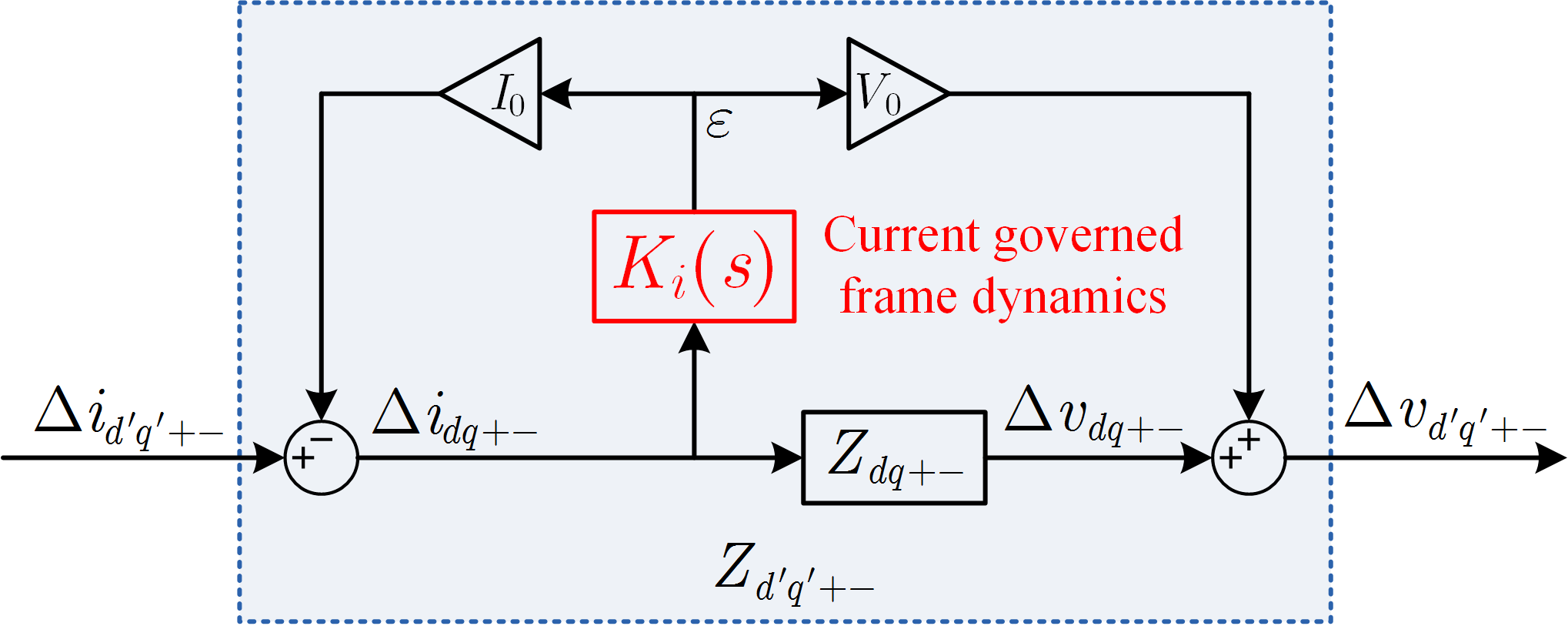}}

\subfloat[]{\includegraphics[scale = 0.95]{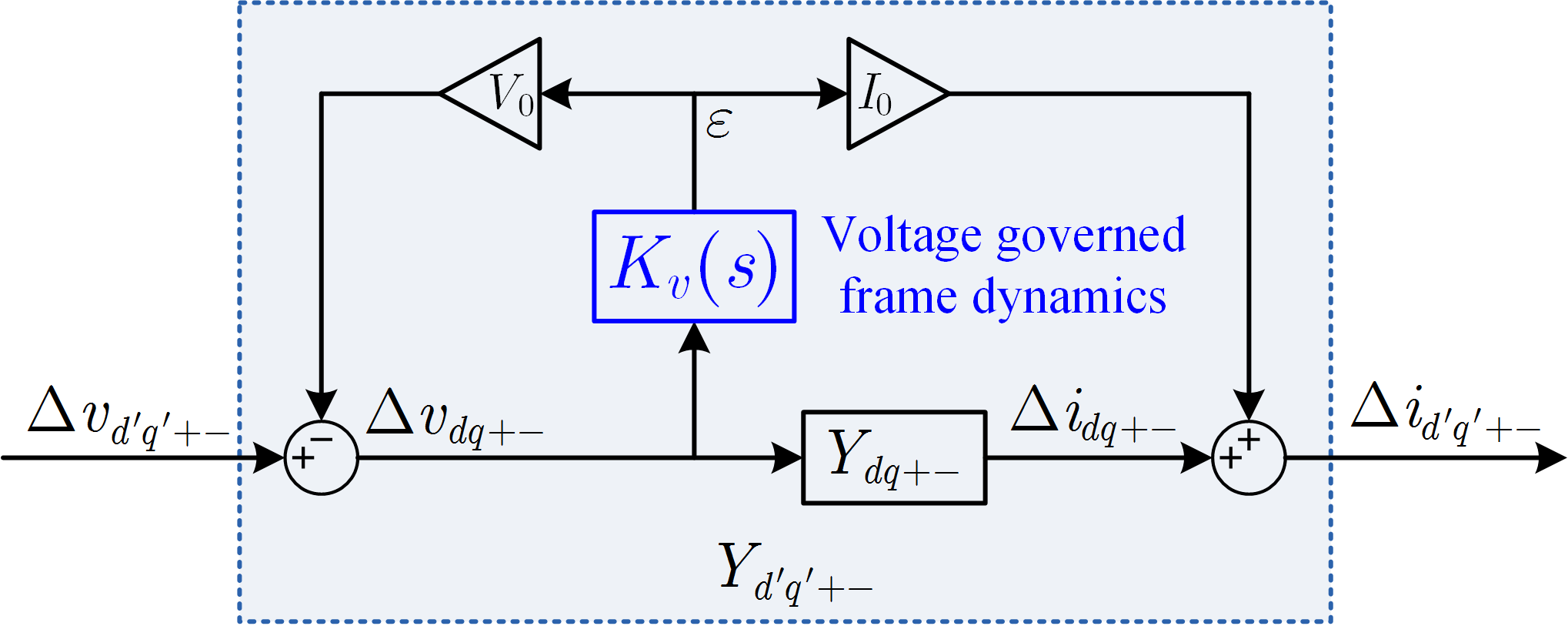}}
\caption{Signal flow in an impedance and admittance model with frame dynamics embedded. (a) Impedance with current-governed frame dynamics. (b) Admittance with voltage-governed frame dynamics.}
\label{fig_loop}
\end{figure}

The second-step transformation (ii) is global but constant, that is, without dynamics. It does require information external to the local frame, namely, the angle $\xi_n$ between the local steady frame and the global steady frame, but $\xi_n$ remains constant in dynamic modeling and can be obtained from power flow analysis \textit{a priori}. The impedance transformation for (ii) is well-understood and the result is stated below without proof:

\begin{equation} \label{eq_G}
\underbrace{
\begin{bmatrix}
Z_{++}  & Z_{+-}   \\
Z_{-+}  & Z_{--}   \\
\end{bmatrix}}_{{Z}_{d^\prime_n q^\prime_n +-}}
\longleftrightarrow
\underbrace{
\begin{bmatrix}
Z_{++} & Z_{+-} e^{j2\xi_n} \\
Z_{-+}e^{-j2\xi_n} & Z_{--} \\
\end{bmatrix}}_{{Z}_{D^\prime Q^\prime +-}}
\end{equation}
in which ${Z}_{d^\prime_n q^\prime_n +-}$ is the impedance of the $n$th machine in its local steady frame, and ${Z}_{D^\prime Q^\prime +-}$ is the corresponding impedance in the global steady frame. The complex signal method is used here to represent the transformation in a concise form \cite{harnefors2007modeling,holtz1995representation}. It is clear from (\ref{eq_G}) that the diagonal entries $Z_{++}$ and $Z_{--}$ are unchanged by the transformation, and the anti-diagonal entries are simply modified by a phase angle $2\xi_n$. No additional dynamics arise from this transformation. 

The first-step transformation (i), on the other hand, is local but dynamic. It carries all frame dynamics and can be defined locally without referring to external information. The impedance transformation for this step is non-trivial and is discussed in detail in the next section.

\section{Frame-Dynamics-Embedding Impedance Transformation}

The impedance transformation from a local swing frame to a local steady frame is the essential step to embed frame dynamics in local impedance models, and is therefore named in this paper as a frame-dynamics-embedding impedance transformation. The transformation law for a generic machine (synchronous generators, induction motors, or power converters) is discussed below.

Following \figref{fig_angle}, let the angle variation between the swing frame $dq$ and the steady frame $d^\prime q^\prime$ be $\epsilon$. Now we seek to determine how the dynamics in $\epsilon$ can be represented in the impedance model. Complex signals are used to represent the current or voltage in $dq$ and $d^\prime q^\prime$ frames and the transformation in between is
\begin{equation}
\label{eq_transform} 
\begin{bmatrix}
u_{d^\prime q^\prime+} \\
u_{d^\prime q^\prime-} \\
\end{bmatrix}
= 
\begin{bmatrix}
e^{j\epsilon} & 0 \\
0 & e^{-j\epsilon}  \\
\end{bmatrix}
\begin{bmatrix}
u_{d q +} \\
u_{d q -} \\
\end{bmatrix}.
\end{equation}
where 
\begin{equation}
\begin{array}{cc}
u_{dq+} = u_d + j u_q,\ u_{d^\prime q^\prime +} = u_{d^\prime} + j u_{q^\prime}\\ 
u_{dq-} = u_d - j u_q,\ u_{d^\prime q^\prime -} = u_{d^\prime} - j u_{q^\prime}
\end{array}
\end{equation}
are complex current or voltage signals. Equation (\ref{eq_transform}) is linearized to see the effect of the frame perturbation
\begin{equation} \label{eq_epsilon}
\begin{bmatrix}
\Delta u_{d^\prime q^\prime+} \\
\Delta u_{d^\prime q^\prime-} \\
\end{bmatrix}
= 
\begin{bmatrix}
\Delta u_{dq+} \\
\Delta u_{dq-} \\
\end{bmatrix}
+
\begin{bmatrix}
ju_{0+} \\
-ju_{0-} \\
\end{bmatrix}
\cdot
\epsilon
\end{equation}
where ${\Delta u}$ is the deviation from the operating point $u_{0}$ and $u_{0}$ is the same for both $u_{d q}$ and $u_{d^\prime q^\prime}$ (because the swing frame and steady frame are aligned in the steady state). Written in a compact form, (\ref{eq_epsilon}) becomes
\begin{equation} \label{eq_epsilon_compact}
\Delta {u}_{d^\prime q^\prime +-} = \Delta {u}_{dq+-} + {U}_{0} \cdot \epsilon
\end{equation}
in which 
\begin{equation}
	\Delta {u}_{dq +-} \!\!=\!\! 
	\begin{bmatrix}
		\!\Delta {u}_{dq +}\! \\
		\!\Delta {u}_{dq -}\! 
	\end{bmatrix}
	\!\!,\,
	\Delta {u}_{d^\prime q^\prime +-} \!\!=\!\!  
	\begin{bmatrix}
		\!\Delta {u}_{d^\prime q^\prime +}\! \\
		\!\Delta {u}_{d^\prime q^\prime -}\! 
	\end{bmatrix}
	\!\!,\,
	{U}_{0} \!\!=\!\!  
	\begin{bmatrix}
		\!ju_{0+}\! \\
		\!-ju_{0-}\! 
	\end{bmatrix}\!\!.
\end{equation}

The frame perturbation $\epsilon$ might be governed by either current (in the case of a synchronous generator) or voltage (in the case of a PLL-controlled converter). Taking the former as an example,
\begin{equation} \label{eq_Kp}
\epsilon(s) = K_i(s) \Delta {i}_{dq+-}(s)
\end{equation}
in which $K_i(s)$ is the transfer function representing the current-governed frame dynamics. Let the impedance in the swing frame be represented as
\begin{equation} \label{eq_Zdq1} 
\Delta v_{dq+-} = Z_{dq+-} \Delta i_{dq+-}
\end{equation}
in which $v$, $i$ and $Z$ denote voltage, current, and impedance respectively. Equation (\ref{eq_Zdq1}) can be transformed to the steady frame using (\ref{eq_Kp}) and (\ref{eq_epsilon_compact}), from which the impedance transformation law is obtained below: 
\begin{equation} \label{eq_Z_trans}
{Z}_{d^\prime q^\prime+-} = ({Z}_{dq+-} + {V}_0 K_i)({I} + {I}_0 K_i)^{-1}
\end{equation}
in which ${Z}_{dq+-}$ and ${Z}_{d^\prime q^\prime+-}$ are the impedance in the swing and steady frame respectively, $I$ is the unit matrix, $V_0 = (jv_{0+} \ ,\ -jv_{0-})^\top$ represents the steady-state voltage, and $I_0 = (ji_{0+} \ ,\ -ji_{0-})^\top$ represents the steady-state current. The admittance transformation law is similar:
\begin{equation} \label{eq_Y_trans}
{Y}_{d^\prime q^\prime+-} = ({Y}_{dq+-} + {I}_0 K_v)({I} + {V}_0 K_v)^{-1}
\end{equation}
in which ${Y}_{dq+-}$ and ${Y}_{d^\prime q^\prime+-}$ are the impedance in the swing and steady frame respectively and $K_v$ represents the voltage-governed frame dynamics. 

The impedance/admittance transformations above can be visualized by the signal flow diagrams in \figref{fig_loop}. Taking \figref{fig_loop}(a) as an example, a current disturbance induces a perturbation (or swing) of the local swing frame, $\epsilon$, which in turn affects the representation of the current and the voltage in the steady frame. As a result, the dynamics implicit in the local swing frame are embedded into the impedance in the steady frame.

\section{Closed-loop Formulation without System Separation}

After the impedance models of all machines have been transformed into the global steady frame, they can be assembled in a modular fashion into impedance model of the whole system. In the conventional use of impedance models, such as in low-voltage dc power network, a system is separated into source-side and load-side. Taking the source voltage or load current as input, and source current or load voltage as output, the whole system then forms a closed-loop transfer function and its stability is evaluated by the ratio of impedances or admittances of the two sides. 

\begin{figure}
\centering
\includegraphics[scale = 1.1]{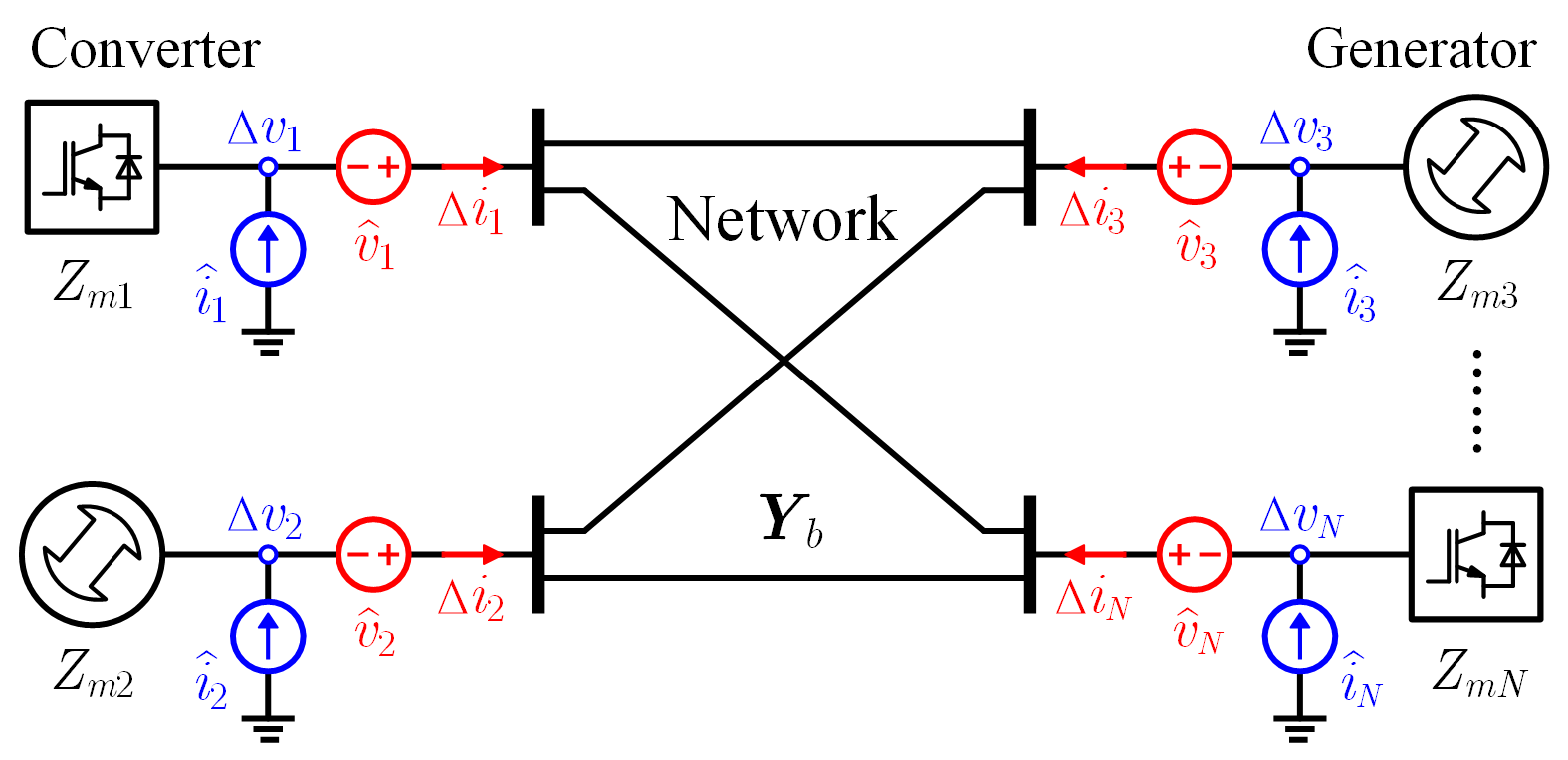}
\caption{Inject virtual input at $\hat{{{v}}}_k$ and $\hat{{{i}}}_k$ and take output at $\Delta{{{i}}}_k$ and $\Delta{{{v}}}_k$ to form the closed loop transfer function.}
\label{fig_inject}
\end{figure}

In a large-scale ac power system, however, there are multiple sources and loads interconnected in a meshed network, which makes the conventional source-load separation very difficult if not impossible. To solve this problem, the formulation in \figref{fig_inject} is developed. Virtual current injection $\hat{i}_k$ ($k = 1,2,\hdots,N$) in parallel with each node, or virtual voltage injection $\hat{v}_k$ in series with each branch is introduced, in which both $\hat{i}_k$ and $\hat{v}_k$ are two-dimensional column vectors in $D^\prime Q^\prime$ frame. The $\hat{i}_k$ and $\hat{v}_k$ such defined do not change the dynamic structure of the system (including the topology of the network and the models of sources and loads) and are very small so as not to affect the operating point around which the system is linearized. 

\begin{figure}
\centering
\includegraphics[scale = 1.1]{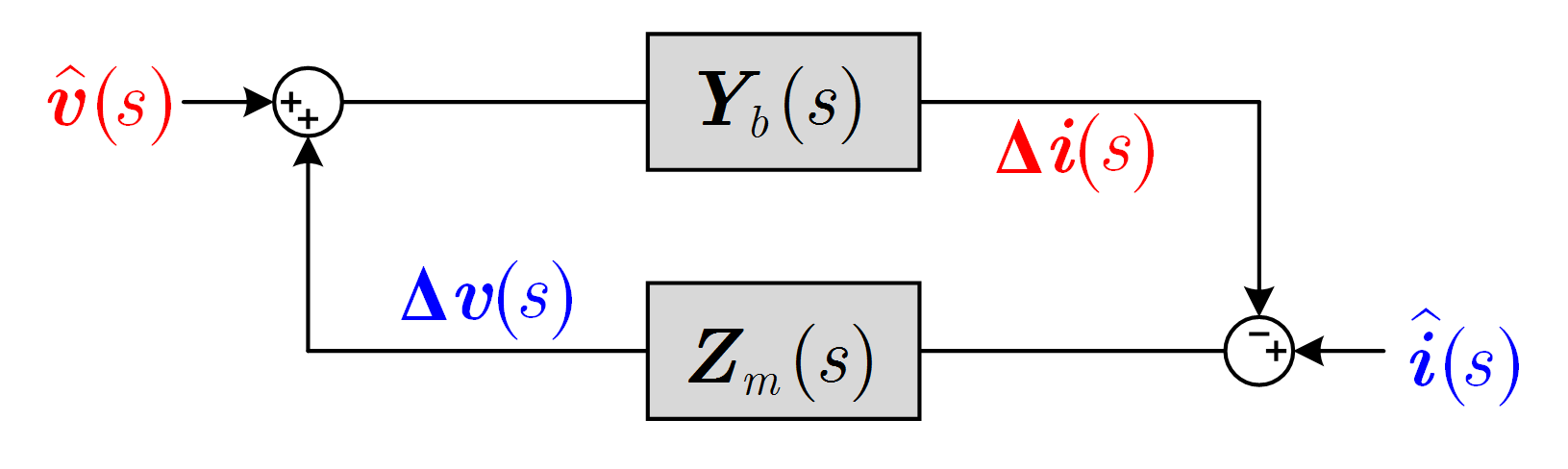}
\caption{Whole-system dynamic modeling by the closed-loop formulation of nodal admittance matrix and machine impedance matrix.}
\label{fig_whole}
\end{figure}

Taking $\hat{i}_k$ or $\hat{v}_k$ as input, and the responding node voltage variation $\Delta {v}_k$ or branch current variation $\Delta {i}_k$ as output, a new closed-loop formulation is obtained as shown in \figref{fig_whole}. In this formulation, the voltages and currents are written concisely as vectors, that is,  
\begin{equation}
\hat{\bm{i}}= \!\!
	\begin{bmatrix}
	\hat{i}_1 \\
	\hat{i}_2 \\
	\vdots \\
	\hat{i}_N \\
	\end{bmatrix}
\!,\ 
\hat{\bm{v}}= \!\! 
	\begin{bmatrix}
	\hat{v}_1 \\
	\hat{v}_2 \\
	\vdots \\
	\hat{v}_N \\
	\end{bmatrix}
\!,\  
\Delta{{\bm{i}}}= \!\! 
	\begin{bmatrix}
	\Delta{i}_1 \\
	\Delta{i}_2 \\
	\vdots \\
	\Delta{i}_N \\
	\end{bmatrix}	
\!,\  
\Delta{{\bm{v}}}= \!\! 
	\begin{bmatrix}
	\Delta{v}_1 \\
	\Delta{v}_2 \\
	\vdots \\
	\Delta{v}_N \\
	\end{bmatrix}	
\!.
\end{equation}
${\bm{Z}_{m}}$ represents the impedances of all machines (including generators, converters and loads) written as a block-diagonal matrix:
\begin{equation}
{\bm{Z}_{m}} = 
\begin{bmatrix}
Z_{{m}1}(s) \\
 & Z_{{m}2}(s) \\
 & & \ddots \\ 
 & & & Z_{{m}N}(s)
\end{bmatrix}
\end{equation}
where $Z_{{m}k}(s)$ is a $2 \times 2$ block in the $D^\prime Q^\prime$ frame representing the dynamic impedance of the $k$-th machine. ${\bm{Y}_{b}}$ is the nodal admittance matrix of the network and represents the dynamics of the transmission lines
\begin{equation}
{\bm{Y}_{b}} = 
\begin{bmatrix}
Y_{{b}11}(s) & Y_{{b}12}(s) & \cdots & Y_{{b}1N}(s) \\
Y_{{b}21}(s) & Y_{{b}22}(s) & \cdots & Y_{{b}2N}(s) \\
     \vdots  &      \vdots  & \vdots &      \cdots  \\
Y_{{b}N1}(s) & Y_{{b}N2}(s) & \cdots & Y_{{b}NN}(s) \\
\end{bmatrix}
\end{equation}
in which the off-diagonal element ${Y}_{bkl}$  ($k \neq l$) represents the branch admittance between the $k$-th and $l$-th nodes and the diagonal element ${Y}_{bkk}$ is the self-admittance at the $k$-th node. In the $D^\prime Q^\prime$ frame with complex signals, ${Y}_{bkl}$ and ${Y}_{bkk}$ are $2\times 2$ matrix blocks in the following forms:
\begin{multline}
{Y}_{bkl}(s) = \\
- 
\begin{bmatrix}
R_{kl} + (s+j\omega_0) L_{kl} \!\! & \\
 & \!\! R_{kl} + (s-j\omega_0) L_{kl}
\end{bmatrix}^{-1}
\end{multline}
\begin{multline}
{Y}_{bkk}(s) = -\sum_{l \neq k}{{Y}_{bkl}}(s) \\
+
\begin{bmatrix}
R^{-1}_{kk} + (s+j\omega_0)C_{kk} \!\!& \\
 & \!\! R^{-1}_{kk} + (s-j\omega_0)C_{kk} 
\end{bmatrix} 
\end{multline} 
where $R_{kl}$ and $L_{kl}$ are the resistance and inductance between the $k$-th and $l$-th nodes, and $R_{kk}$ and $C_{kk}$ are the resistance and capacitance connected at the $k$-th node. It is worth noting that ${\bm{Y}_{b}}(s)$ such defined is different to the conventional static nodal admittance matrix since the dynamics of the network are included in the frequency domain $s$.

According to \figref{fig_inject}, the machine terminal voltage $\Delta{\bm{v}}$ adds to the virtual voltage $\hat{\bm{v}}$, which are jointly applied on the network nodal admittance matrix ${\bm{Y}_{b}}$ and thereby generates the nodal current $\Delta{\bm{i}}$. The nodal current $\Delta{\bm{i}}$ is then deduced (because of the sign convention marked in \figref{fig_inject}) from the virtual current $\hat{\bm{i}}$ and injected to the machine impedance matrix ${\bm{Z}_{m}}$ to generate the machine voltage $\Delta{\bm{v}}$. This feedback relationship forms the equivalent closed-loop diagram in \figref{fig_whole}. Accordingly, the stability of the whole system is determined by the interaction between ${\bm{Z}_{m}}$ and ${\bm{Y}_{b}}$ rather than the interaction between source-side and load-side, which thereby avoids the source-load separation. 

Under this formulation, the whole-system impedance $\hat{{\bm{Z}}}$ is defined as the transfer function matrix from $\hat{\bm{i}}$ to $\Delta{\bm{v}}$
\begin{equation}
\Delta{\bm{v}} = \hat{{\bm{Z}}} \cdot \hat{\bm{i}} \ \big{|}_{\hat{\bm{v}}=\bm{0}} ,\ \hat{{\bm{Z}}} = {\bm{Z}_{m}} ({\bm{I}} + {\bm{Y}_{b}}{\bm{Z}_{m}})^{-1}
\end{equation}
and similarly, the whole-system admittance $\hat{{\bm{Y}}}$ is defined as the transfer function matrix from $\hat{\bm{v}}$ to $\Delta{\bm{i}}$
\begin{equation}
\label{eq_Yhat}
\Delta{\bm{i}} = \hat{{\bm{Y}}} \cdot \hat{\bm{v}} \ \big{|}_{\hat{\bm{i}}=\bm{0}} ,\ \hat{{\bm{Y}}} = ({\bm{I}} + {\bm{Y}_{b}}{\bm{Z}_{m}})^{-1}{\bm{Y}_{b}}.
\end{equation}

\figref{fig_inject} and \figref{fig_whole} can be converted to the dual form where $\bm{Z}_{m}$ and $\bm{Y}_{b}$ are replaced by $\bm{Y}_{m}$ and $\bm{Z}_{b}$ respectively, and
\begin{equation}
\bm{Y}_{m} = \bm{Z}_{m}^{-1} ,\ \bm{Z}_{b} = \bm{Y}_{b}^{-1}.
\end{equation}
Accordingly, the whole-system impedance and admittance are represented by
\begin{equation}
\hat{\bm{Y}} = \bm{Y}_{m}(\bm{I} + \bm{Z}_{b}\bm{Y}_{m})^{-1} ,\ \hat{\bm{Z}} = (\bm{I} + \bm{Z}_{b}\bm{Y}_{m})^{-1}\bm{Z}_{b}. 
\end{equation}
This formulation is more convenient when the machine admittance $\bm{Y}_{m}$ can be more readily obtained than the machine impedance $\bm{Z}_{m}$.

{In the conventional impedance-based analysis, Nyquist criterion is usually used to evaluate the closed-loop stability of the system from the open-loop impedance curves. This approach is not suitable for the proposed network separation method as both ${\bm{Z}}_m$ and ${\bm{Y}}_b$ are high-dimensional matrices. It is possible to use the generalized Nyquist criterion to handle these matrices but this leads to loss of connection with the original system so the stability evaluation becomes very hard to interpret. To solve this problem, we directly use the closed-loop impedance and admittance, $\hat{{\bm{Z}}}(s)$ and $\hat{{\bm{Y}}}(s)$, to evaluate stability. $\hat{{\bm{Z}}}(s)$ and $\hat{{\bm{Y}}}(s)$ are matrices in which each elements represent the impedance and admittance seen at the different nodes in the complete system. All elements share the same poles since the network is interconnected and oscillatory signals will propagate to all nodes in the end. Therefore, any element in $\hat{{\bm{Z}}}(s)$ or $\hat{{\bm{Y}}}(s)$ can be used to evaluate the stability of the whole system via evaluating the poles of the corresponding transfer function. However, the poles seen at various elements may have different magnitudes but this provides an indication of the relative participation of different nodes in a particular mode, as discussed in the following section.}

\section{Composite Generator-Converter Grid}

In this section, the proposed methodology will be applied to whole-system modeling of a composite converter-generator grid. Using the impedance transformation described in the preceding section, we first find the frame-dynamics-embedded impedance for both synchronous generators and PLL-controlled converters in local steady frames, and then interconnect them to create a whole-system model. From this whole-system model, an interesting result is illustrate with concerning the interactions between generator rotors and converter control in a low-inertia grid.

\subsection{Synchronous Generator}
The state equations of a synchronous generator in the local swing frame $dq$ is
\begin{equation}
\begin{array}{l}
\dot{\psi}_{d} = v_{d} - R i_{d} + \omega \psi_{q} \\
\dot{\psi}_{q} = v_{q} - R i_{q} - \omega \psi_{d} \\
\dot{\omega} = (T_{e} - T_{m} - D{\omega})/J\\
\dot{\theta} = \omega
\end{array}
\end{equation}
in which
\begin{equation}
\psi_{d} = L i_{d} ,\ \psi_{q} = L i_{q} - \psi_f,\ T_{e} = \psi_f i_{d}.
\end{equation}
The symbols are defined as $\psi$, $v$ and $i$ being the flux-linkage, voltage, and current respectively, $\psi_f$ is the field flux-linkage, $\omega$ and $\theta$ are the rotor speed and angle, $T_{e}$ and $T_{m}$ are the electrical and mechanical torque, $R$ and $L$ are the stator resistance and inductance respectively, $D$ is the damping torque coefficient, and $J$ is the rotor inertia. The constant flux-linkage model ($\psi_f = \text{constant}$) is used which combines the total armature reaction in $L$, and $T_{m}$ is assumed to be constant on the basis that the prime-mover's speed governor is slow compared to the fast transients under consideration \cite{kundur1994power}. A single pole-pair with no saliency is considered to simplify the model without losing the essential properties. All variables are in motor convention and the $q$ axis is aligned to the field flux so that $i_d$ and $i_q$ represent active and reactive current seen by the generator respectively.

Linearizing the state equation and using Laplace transformation, the impedance in the local swing frame is obtained
\begin{equation} \label{eq_ss_sg}
Z_{d q} 
\approx
\begin{bmatrix}
Z_L(s)      &  -\omega_0 L    \\
\omega_0 L  &   Z_L(s)        \\
\end{bmatrix}
\end{equation}
or equivalently in the complex-signal form
\begin{equation} \label{eq_Zp}
Z_{d q+-} 
\approx
\begin{bmatrix}
Z_L(s_1) &                 \\
    	   & Z_L(s_{-1})     \\					
\end{bmatrix}
\end{equation}
in which $Z_L(s) = sL + R$, $s_{1} = s + j\omega_0$, and $s_{-1} = s - j\omega_0$. $Z_{d q+-}$ is transformed into the steady frame next. The swing ${\epsilon}$ is determined by the perturbation of the rotor angle ${\Delta \theta}$, which in turn is governed by electric torque $\Delta T_{e}$ proportional to current $\Delta i_{d}$. That is,
\begin{equation}
\epsilon (s) = \Delta \theta (s) = \frac{\Delta i_{d}(s)}{sH(s)} = \frac{\Delta i_{d q +}(s) + \Delta i_{d q -}(s)}{2sH(s)}
\end{equation}
where $H(s) = {(Js+D)}/{\psi_f}$, from which follows the current-governed frame dynamics
\begin{equation}
{K_i} = \frac{1}{2sH(s)} 
\begin{bmatrix}
1 & 1
\end{bmatrix}.
\end{equation}
The steady-frame impedance can then be obtained using the transformation law (\ref{eq_Z_trans}):
\begin{equation} \label{eq_Zp_sg}
{Z}_{d^\prime q^\prime +-} = 
\underbrace{
\begin{bmatrix}
Z_L(s_{1}) \!\!&\!\!             \\
         \!\!&\!\! Z_L(s_{-1})   \\
\end{bmatrix}
}_\text{flux dynamics}
+
\underbrace{
\frac{1}{M(s)}
\begin{bmatrix}
s_{1}      \!\!&\!\! s_{1}   \\
s_{-1}     \!\!&\!\! s_{-1}  \\
\end{bmatrix}
}_\text{frame dynamics}
\end{equation}
in which 
\begin{equation} \label{eq_M_sg}
M(s) = \frac{2}{\psi_f^{2}}(Js^2 + Ds - i_{q0}\psi_f).
\end{equation}
It is clear that ${Z}_{d^\prime q^\prime +-}$ contains two parts. The first part is the same as ${Z}_{dq +-}$ and represents the flux dynamics of the windings. The second part only appears in the steady frame and represents frame dynamics of the rotor, which manifests the effect of frame-dynamics-embedding transformation. It is worth noting that the second part may have right-half-plane (RHP) poles in $M(s)$, which leads to very interesting implications on the stability of rotor-control interaction, as explained in the subsection after next.

\subsection{PLL-Controlled Converter}
The steady-frame impedance for a PLL-controlled grid-connected converter can be found in a similar way to that of a synchronous generator, except that the frame dynamics of a PLL is governed by voltage rather than current. The PLL takes voltage $v_q$ to estimate the angle difference and changes the internal rotating frequency and angle accordingly, so the voltage governed frame dynamics can be represented as:
\begin{equation}
{K_v} = \frac{1}{2j H_\text{PLL}} 
\begin{bmatrix}
1 & -1
\end{bmatrix}
\end{equation}
where
\begin{equation}
H_\text{PLL} = v_{d0} + \frac{v_{d0} s^2}{k_{p} s + k_{i}}
\end{equation}
$k_{p}$ and $k_{i}$ are the proportional and integral gain of PLL, and $v_{d0}$ is the operating point of $v_d$. Based on the $K_v$ above, we invoke the admittance transformation in (\ref{eq_Y_trans}) to embed the frame dynamics, and invert the admittance matrix to obtain the corresponding impedance. This process is trivial but tedious so the Matlab symbolic calculation is used to find the expression, and the Matlab scripts can be found at \cite{toolbox}.
 
\subsection{Generator-Converter Interaction}

\begin{figure}
\centering
\includegraphics[scale = 1.0]{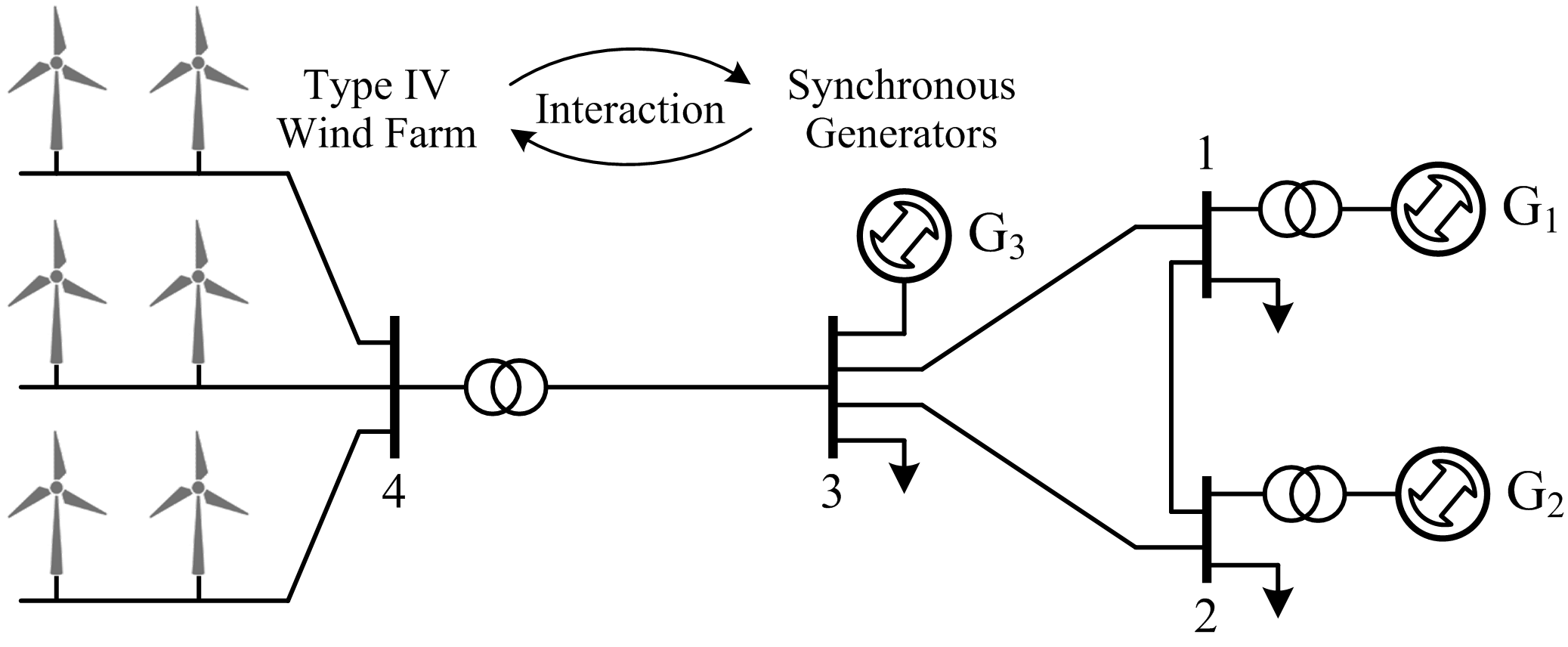}
\caption{Configuration of the investigated offshore wind farm connected to a low-inertia transmission system.}
\label{fig_windfarm}
\end{figure}

Now the frame-dynamics-embedded impedance models for different machines in the global steady frame is connected for whole-system stability analysis. The stability of generator-converter interaction in a low-inertia system is focused, which shows the advantage of the proposed methodology and is not widely studied in literature. As a specific example, a wind farm connected to the transmission system shown in \figref{fig_windfarm} is investigated. The system has a meshed topology with three synchronous generators ($\text{G}_1$, $\text{G}_2$ and $\text{G}_3$) and a Type-IV wind farm. The turbine-side converters are modeled as constant power sources injecting current into the dc-link capacitors, and the grid-side converters use the standard current vector control with the PLL and dc-link control setting the reference current angle and magnitude. $\text{G}_3$ is configured to have lower power inertia (due to lower power rating) than those of $\text{G}_1$ and $\text{G}_2$. Detailed layouts and parameters are available at \cite{toolbox}.

\begin{figure}
\centering
\subfloat[]{\includegraphics[scale = 0.63]{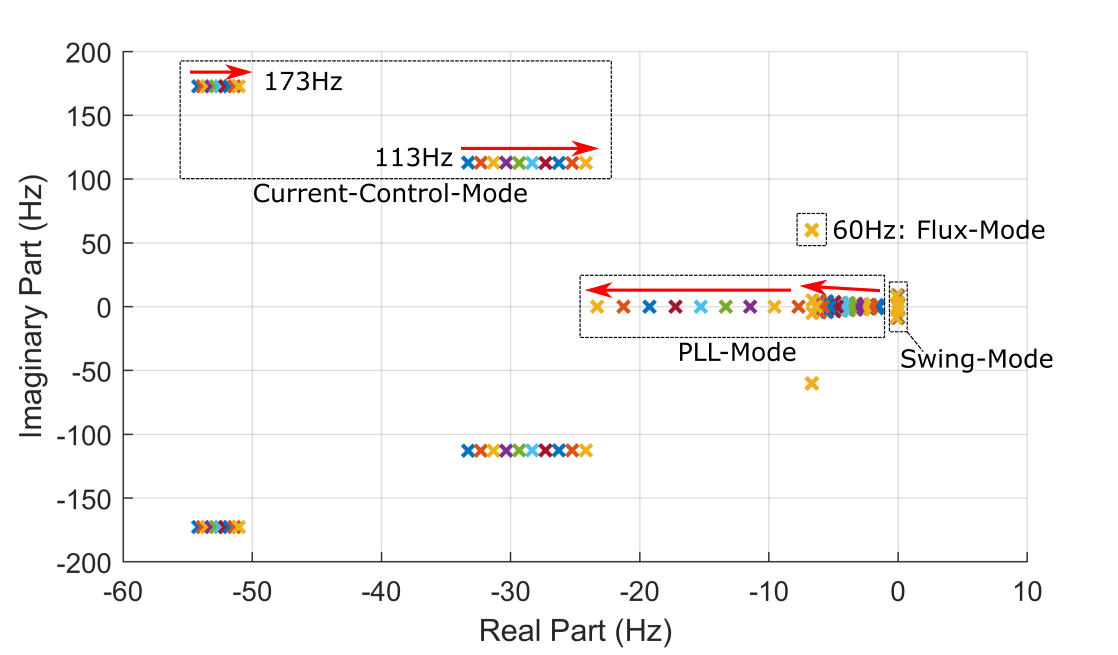}}

\subfloat[]{\includegraphics[scale = 0.63]{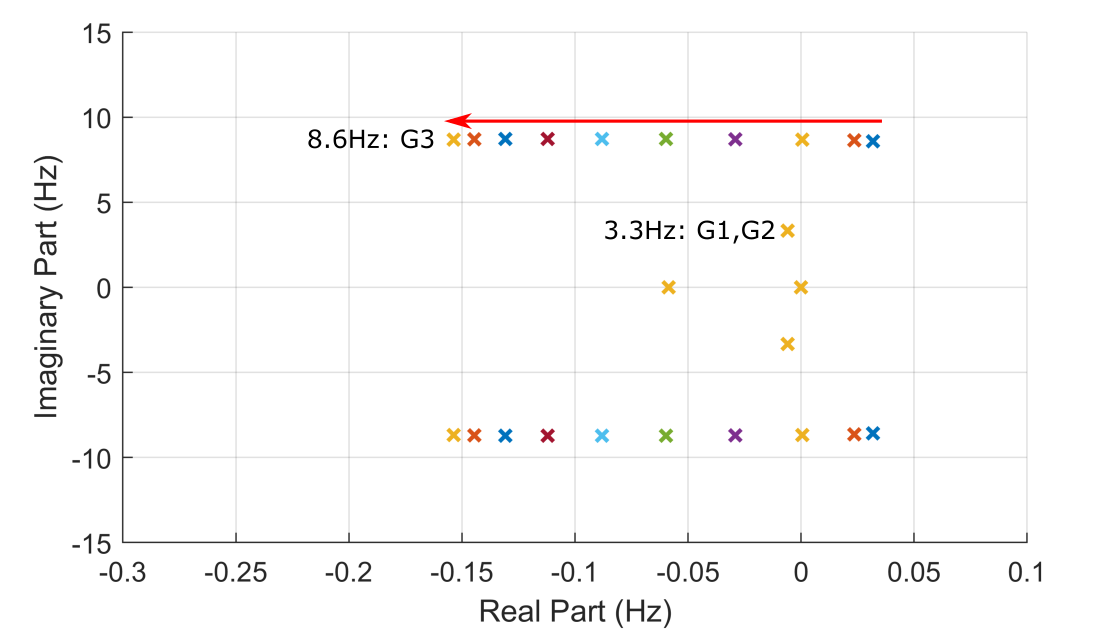}}
\caption{Pole map of $\hat{\bm{Y}}(s)$ when PLL and dc-link control bandwidth sweep from 5Hz to 20Hz. (a) Poles within 200Hz. (b) Zoom into low frequency swing-mode poles. The red arrows indicates that the PLL and dc-link control bandwidth increases. Along the arrows the associated PLL-mode poles move towards higher frequency. Correspondingly, the current-control-mode poles become under-damped whereas the swing-mode poles at 8.6Hz become well-damped due to cross-mode coupling.}
\label{fig_pole}
\end{figure}

\begin{figure}
\centering
\includegraphics[scale = 0.6]{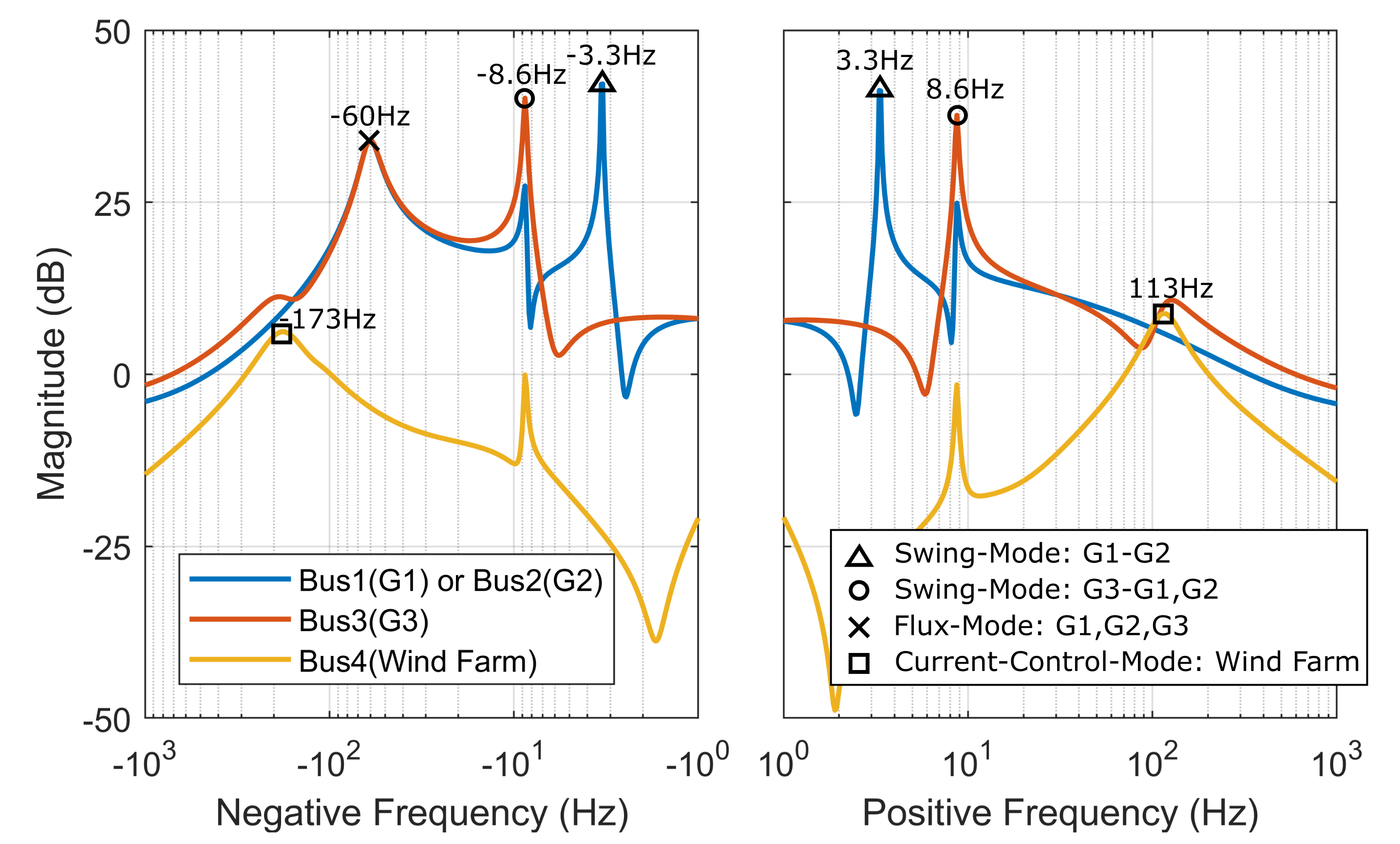}
\caption{System admittance spectrum seen at different buses. The poles in \figref{fig_pole} are reflected as peaks in the admittance curves but with different magnitude at different buses. This further indicates the participation of each bus (and the generators and converters connected to the bus) in a particular mode. The swing mode at 3.3Hz induces high peaks in the admittances at Bus1 and Bus2, indicating that this mode is related to the swing between $\text{G}_1$ and $\text{G}_2$. The swing mode at 8.6Hz induces a high peak in the admittance at Bus3 and moderate peaks at Bus1 and Bus2, indicating that this mode is related to the swing of $\text{G}_3$ against $\text{G}_1$ and $\text{G}_2$. Note that the admittances at Bus1 and Bus2 have an identical curve in the figure as they have the same parameters.} 
\label{fig_admittance_sys}
\end{figure}

\begin{figure*}
\centering
\subfloat[]{\includegraphics[scale = 0.17]{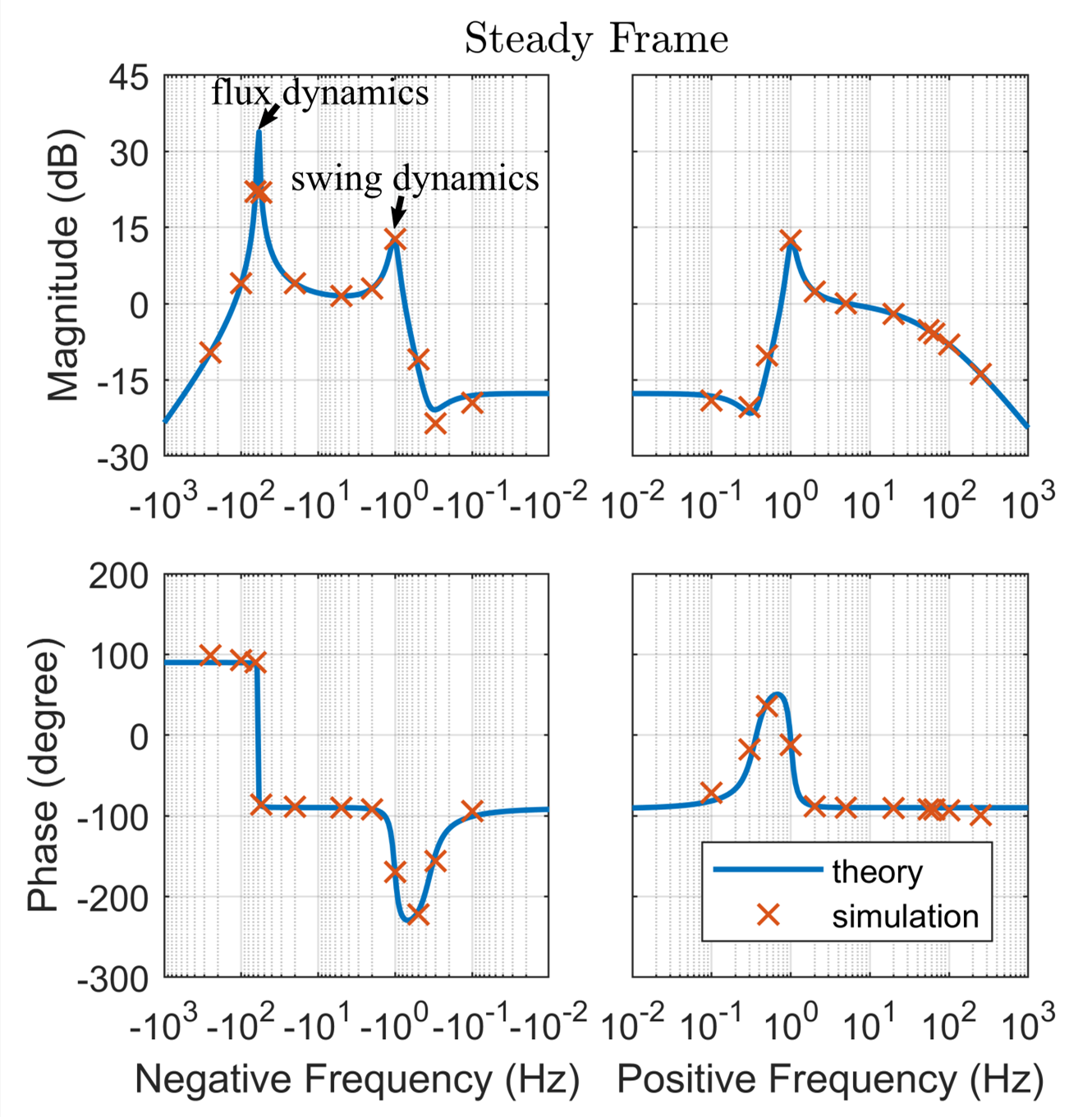}} \ 
\subfloat[]{\includegraphics[scale = 0.17]{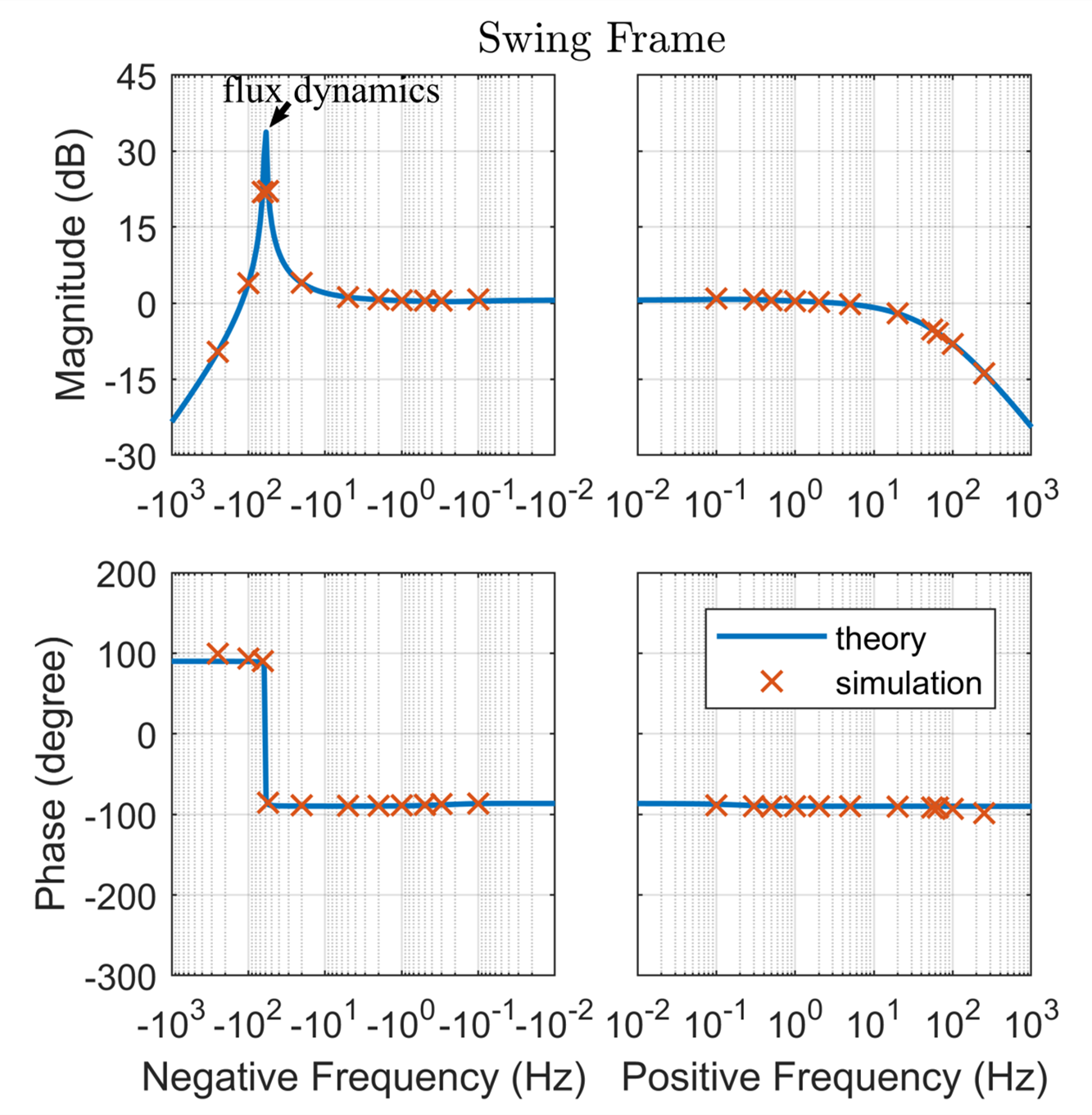}} \ 
\subfloat[]{\includegraphics[scale = 0.17]{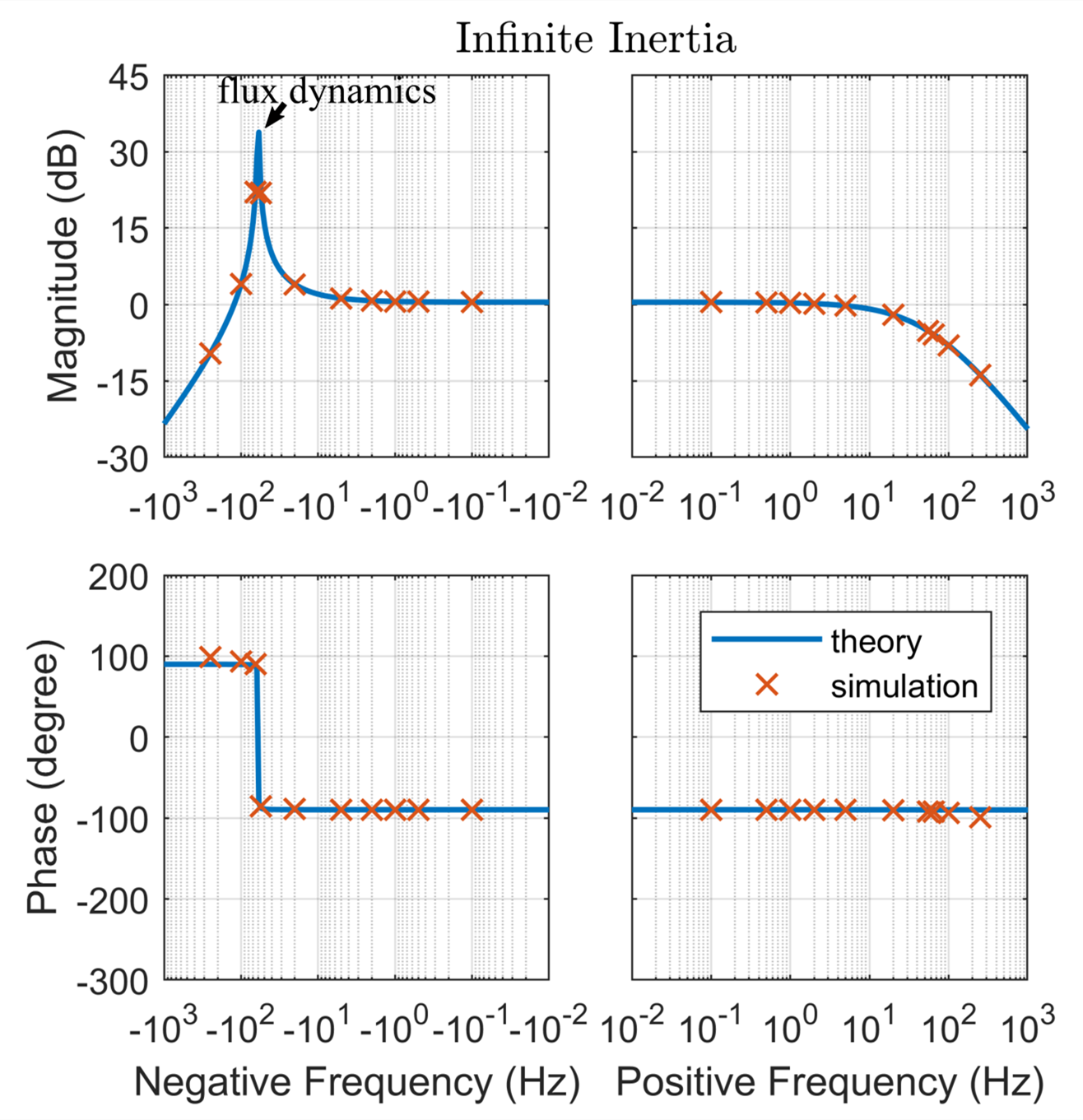}} \ 
\caption{Admittance spectrum of the synchronous generator in various reference frames. (a) Steady frame. (b) Swing frame. (c) In both the steady frame and the swing frame but with constant rotor speed (infinite inertia).}
\label{fig_Ydq}
\end{figure*}

\begin{figure*}
\centering
\subfloat[]{\includegraphics[scale = 0.45]{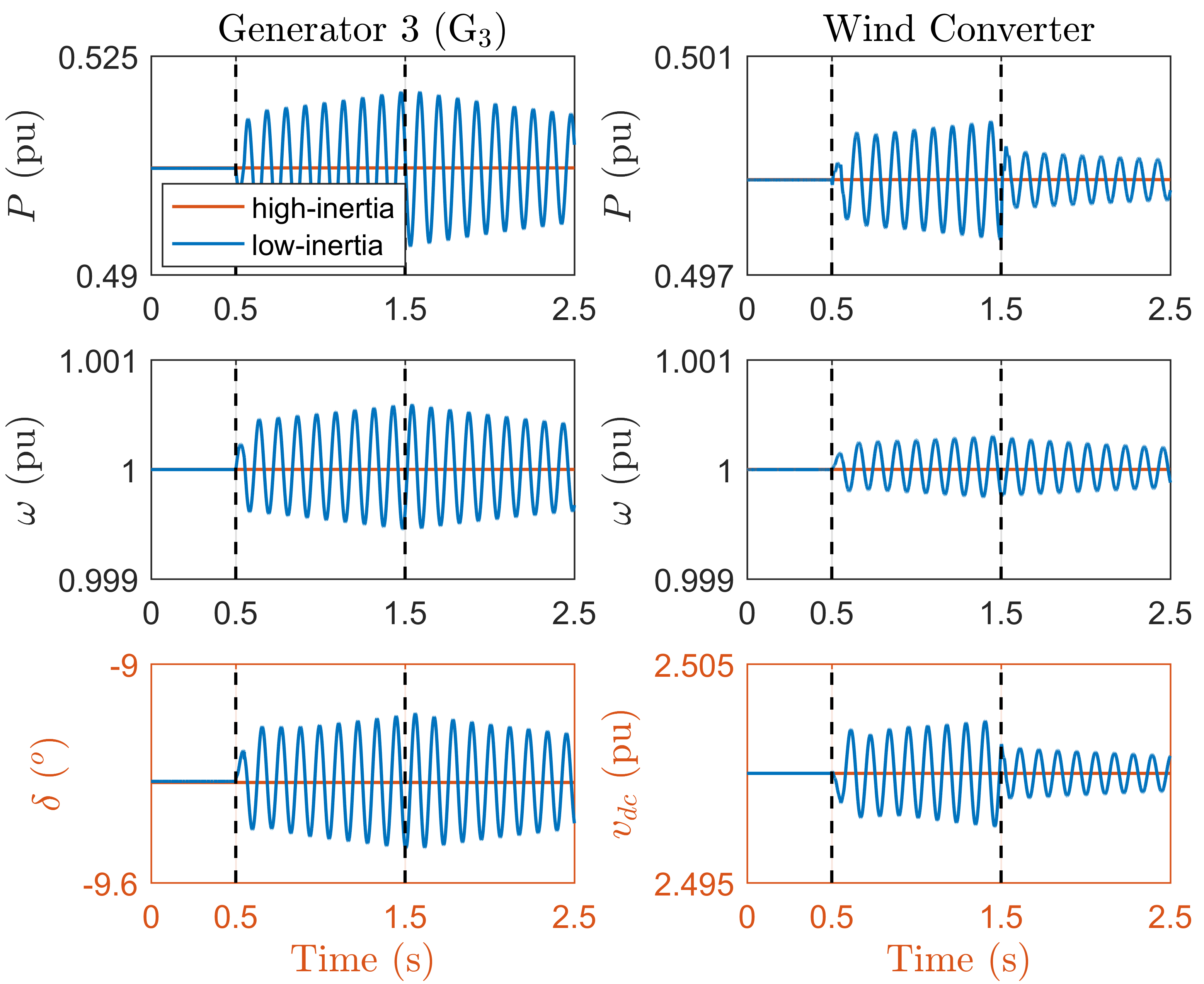}} \ 
\subfloat[]{\includegraphics[scale = 0.45]{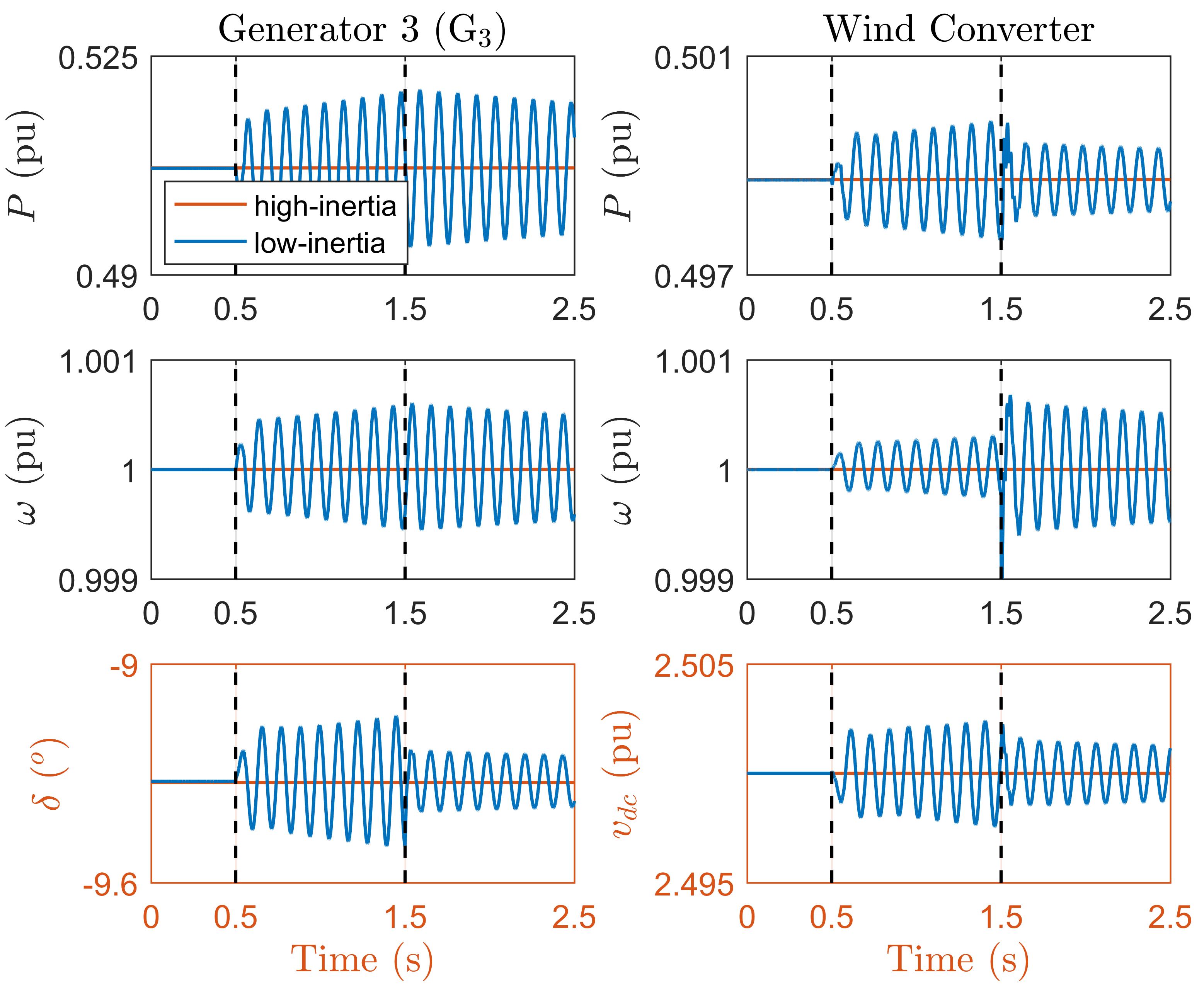}} \ 
\subfloat[]{\includegraphics[scale = 0.45]{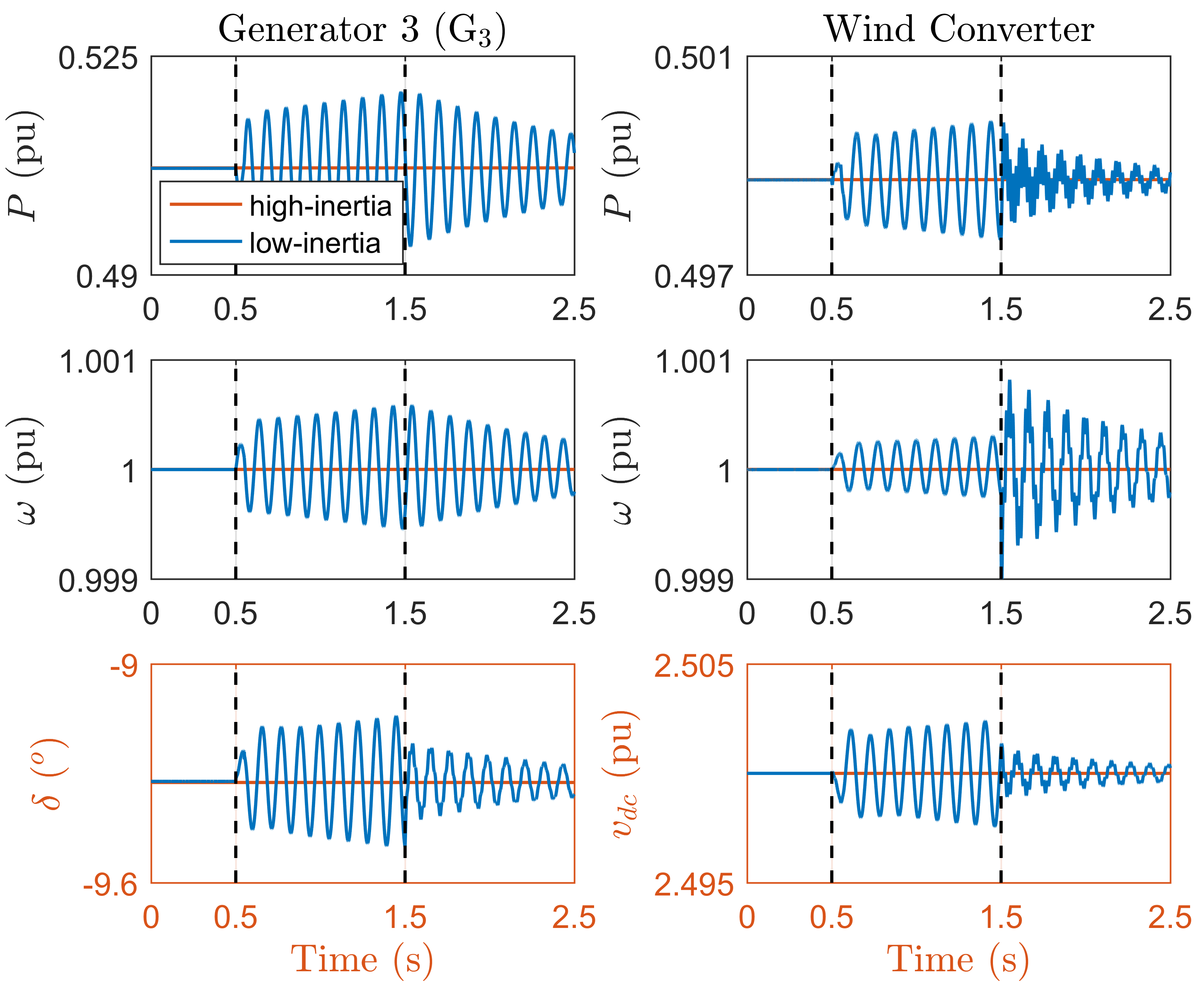}} \ 
\caption{System dynamic response with reducing the bandwidths of dc-link control loop and PLL to 5Hz at 0.5s, and then increasing one or both of them back to 20Hz at 1.5s. (a) Increase the dc-link control bandwidth. (b) Increase the PLL bandwidth. (c) Increase both bandwidths.} 
\label{fig_sim}
\end{figure*}

The impedances of the wind farm and synchronous generators in their local swing frame are modeled first, and then transformed to the local steady frame to embed the frame dynamics, and finally aligned to the global frame to get the machine impedance matrix ${\bm{Z}}_m$. The whole-system admittance $\hat{\bm{Y}}$ is calculated from ${\bm{Z}}_m$ using the closed-loop formulation in (\ref{eq_Yhat}). The poles of $\hat{\bm{Y}}$ indicate the whole-system dynamics and stability and are calculated for a range of converter control parameters as displayed in \figref{fig_pole}. These poles appear in four groups and indicate different oscillation modes in the whole system: 1) the swing-mode poles at low frequency, which are related to the rotor angle swing among the generators; 2) the PLL-mode poles in the mid frequency, which are related to the PLL and dc-link control (PLL and dc-link control are tightly coupled and are therefore categorised as the same mode); 3) the flux-mode poles at the fundamental frequency (60Hz), which are related to the flux transient on winding and line inductances; and 4) the current-control-mode poles at high frequency, which are related to the current control of wind converters.

The zoom-in plot in \figref{fig_pole}(b) further shows that the swing-mode poles contain two sub-groups. The ones at 3.3Hz are related to the swing between $\text{G}_1$ and $\text{G}_2$, and the ones at 8.6Hz are related to $\text{G}_3$ swinging against $\text{G}_1 \text{-} \text{G}_2$ together. This relevance can be seen from \figref{fig_admittance_sys} where the admittances seen at various buses (that is, different entries of the $\hat{\bm{Y}}$ matrix) are plotted. The relative magnitude of the peaks in the impedance curves indicates the participation of different buses in an oscillation mode of a particular frequency. Both positive and negative frequencies are shown in the admittance spectrum, which correspond to the positive and negative sequence signals (or equivalently clockwise and anticlockwise space vectors) respectively, according to the complex signal modeling theory \cite{harnefors2007modeling,holtz1995representation}.

We can further investigate the coupling between different modes. Such coupling may exist between modes with adjacent frequency ranges. From \figref{fig_pole}, we can clearly see two types of cross-mode coupling: 1) PLL-current coupling between the current-control-mode (high frequency) and the PLL-mode, and 2) PLL-swing coupling between the PLL-mode (mid frequency) and the swing-mode (low frequency). The current-control-mode becomes under-damped when the PLL-mode increases in frequency, as marked by the red arrows in \figref{fig_pole}. On the other hand, the swing-mode poles for $\text{G}_3$ become under-damped and even unstable when the PLL-mode decreases in frequency. Therefore, the PLL-mode and the associated PLL and dc-link control bandwidths should be constrained in a specific frequency range.

The PLL-current coupling is well-understood in literature \cite{gu2018reduced} but the PLL-swing coupling is newly observed in this paper. This new PLL-swing coupling can be interpreted by the proposed frame-dynamics-embedded impedance model. If the PLL or the dc-link control of the wind converters are faster (have higher bandwidth) than the rotor swing dynamics, the wind farm follows the grid perfectly and acts as a constant power source. This translates into constant electric torque on the rotors which does not affect the swing dynamics. On the other hand, if the dc-link control and the PLL are slower (have lower bandwidth) than the swing dynamics, the converters are seen as a current source with constant magnitude (due to slow dc-link control) and constant phase (due to slow PLL). The resulted electric torque is affected by the angle difference between converter current and generator rotor which implies that the converter control interacts with rotor swing dynamics. This interaction can be reflected by the impedance of the synchronous generators in the steady frame, since the rotor dynamics is essentially equivalent to the frame dynamics which is embedded in this impedance. According to (\ref{eq_Zp_sg}) and (\ref{eq_M_sg}), the impedance has RHP poles in its second term, which implies instability under constant current injection from wind converters. This leads to the conclusion that the rotor inertia and the corresponding swing frequency sets the lower bound of the bandwidth of PLL and dc-link control of converters, which is the new insight arising from the impedance-based whole-system model.

\section{Simulation Results}

The importance of embedding frame dynamics in impedance models is demonstrated first. A time-domain impedance measurement was conducted on a synchronous generator in both the local swing frame and local steady frame, and the results are compared with the theoretical models in \figref{fig_Ydq}. Admittances instead of impedances are presented here since impedances have right-half-plane poles (see equation (\ref{eq_Zp_sg}) and (\ref{eq_M_sg}) in Section V) and hence are not directly measurable. Impedences can be reconstructed indirectly from the admittance spectrum and the right-half-plane poles can thus be inferred via the method presented in \cite{liao2019impedance}. As in \figref{fig_admittance_sys}, both positive and negative frequencies are shown in the admittance spectrum. In both the steady-frame and swing-frame models in \figref{fig_Ydq} (a) and (b), there is a resonant peak at the fundamental frequency (-60Hz) which represents the flux dynamics $(s + j\omega_0) L$ of the winding and line inductances. In the steady-frame model there is a further resonant peak at $\pm$1Hz, which arises from the frame dynamics and represents the swing mode of the rotor. This resonant peak is missing in the swing-frame model and if this model was used in whole-system analysis then important interactions would be missed. The absence of the swing mode in the swing frame and presence in the steady frame demonstrates that the transformation from swing frame to steady frame embeds frame dynamics into the model. The fact that the 1Hz mode has a proper physical origin is illustrated by setting the rotor inertia $J$ to $\infty$, which then causes the frame dynamics to disappears and the steady- and swing-frame admittances become the same, as shown in \figref{fig_Ydq}(c). This further confirms that the extra resonant peak in \figref{fig_Ydq}(c) does come from the embedded frame dynamics. The results from measurement by signal injection in the time-domain simulation agree with the theoretical models in all three cases except for minor phase errors (which are believed to be caused by the time delays in the discrete-time sampling), showing the very high accuracy of the models.

The key results of whole-system stability analysis based on impedance models are verified next. The system in \figref{fig_windfarm} was simulated via electromagnetic transient (EMT) simulation in Matlab/Simulink. $\text{G}_3$ was configured to have low inertia and the whole system was made temporarily unstable by adjusting the controller speed (bandwidths) of the wind power converters. The simulation results are shown in \figref{fig_sim} and will be compared with the theoretic prediction from the previous section. At 0.5s, the bandwidths of PLL and dc-link control are reduced from 20Hz to 5Hz and returned to their original values at 1.5s separately in plots (a) (b) and (c). Negatively damped oscillation is seen to begin at 0.5s in all signals and the oscillations become positively damped at 1.5s with control bandwidths (either PLL, dc-link control, or together) increased back to 20Hz. These results illustrate the unstable behavior predicted in \figref{fig_pole} and serve to validate the impedance-based modeling methodology via frame-dynamics-embedding transformation and whole-system closed-loop formulation.

Notably, $v_{dc}$ (the dc-link voltage of wind converter) is re-stabilized faster in (a) because of the increased bandwidth of dc-link control; and $\delta$ (the angle difference between the PLL of the wind converters and the rotor of the synchronous generator) is re-stabilized faster in (b) due to the increased bandwidth of PLL. (c) shows the best system stability performance in damping the PLL-swing oscillation, but induces new under-damped modes with a relatively high frequency due to PLL-current coupling. As a result, we come to the final conclusion that either PLL or dc-link control should be fast enough to avoid unstable rotor swing in a low-inertia grid, provided that the current-control-mode is not excited. In other words, PLL-swing and PLL-current interaction set the lower and upper bounds of the PLL and the dc-link control bandwidths respectively for converters in a composite grid.

\section{Conclusions}
This paper establishes a systematic methodology for dynamic modeling of a composite generator-converter power system via impedance models alone. This methodology overcomes the fundamental difficulty of dynamic frame alignment in impedance modeling by embedding the frame dynamics in impedance itself via the transformation between a local steady frame and a swing frame for each generator or power converter. Individual generator or converter models can then be readily combined into a whole-system model even for meshed networks via the proposed closed-loop formulation without network separation. The modeling process and its application to example composite grids leads to new insights into, and possible solutions for, the unstable interactions between converter control and generator rotors in a low inertia grids. This methodology is generally applicable to ac power systems with various topologies and combinations of machines, and can be extended to analyse hybrid ac-dc power systems as well.

\ifCLASSOPTIONcaptionsoff
  \newpage
\fi

\bibliographystyle{IEEEtran}
\bibliography{References}

\begin{IEEEbiography}[{\includegraphics[width=1in,height=1.25in,clip,keepaspectratio]{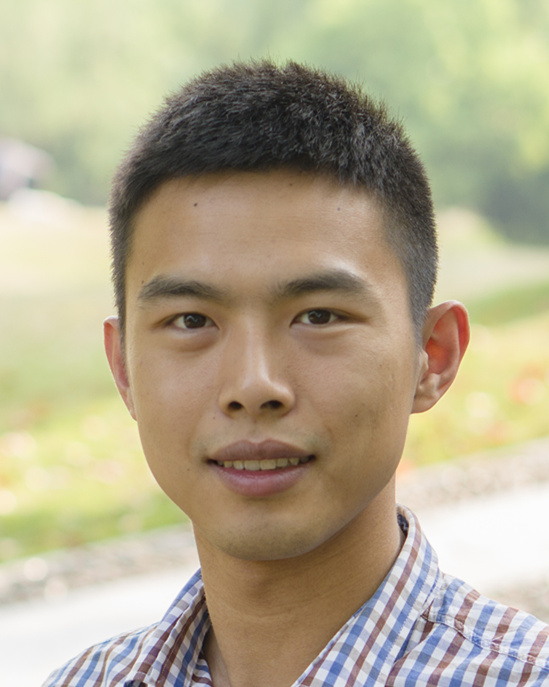}}]{Yunjie Gu}
(M'18-SM'20) received the B.Sc. and the Ph.D. degree in Electrical Engineering from Zhejiang University, Hangzhou, China, in 2010 and 2015 respectively. He was a Consulting Engineer at General Electric Global Research Centre, Shanghai, from 2015 to 2016, and is now an EPSRC-funded Innovation Fellow at Imperial College London (award EP/S000909/1). His research interests include power system control and stability, and the application of power electronics to power systems.
\end{IEEEbiography}

\begin{IEEEbiography}[{\includegraphics[width=1in,height=1.25in,clip,keepaspectratio]{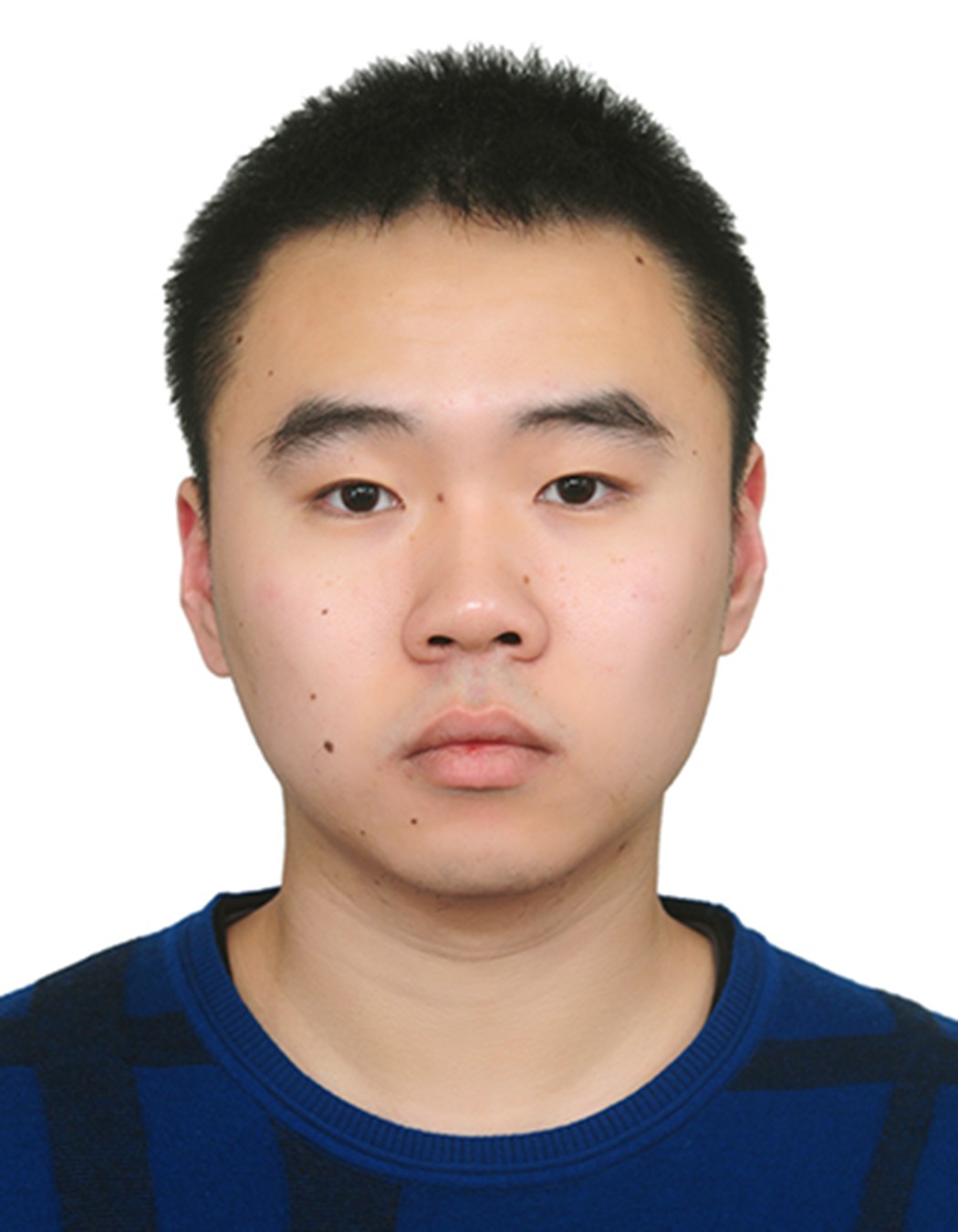}}]{Yitong Li}
(S'17) received the B.Eng degrees from Huazhong University of Science and Technology, China, and the University of Birmingham, UK, in 2015. He received his M.Sc degree in Future Power Networks from Imperial College London, UK, in 2016, where he is currently pursuing his Ph.D. degree. His current research interests include control techniques of power electronic converters and power system stability analysis.
\end{IEEEbiography}

\begin{IEEEbiography}[{\includegraphics[width=1in,height=1.25in,clip,keepaspectratio]{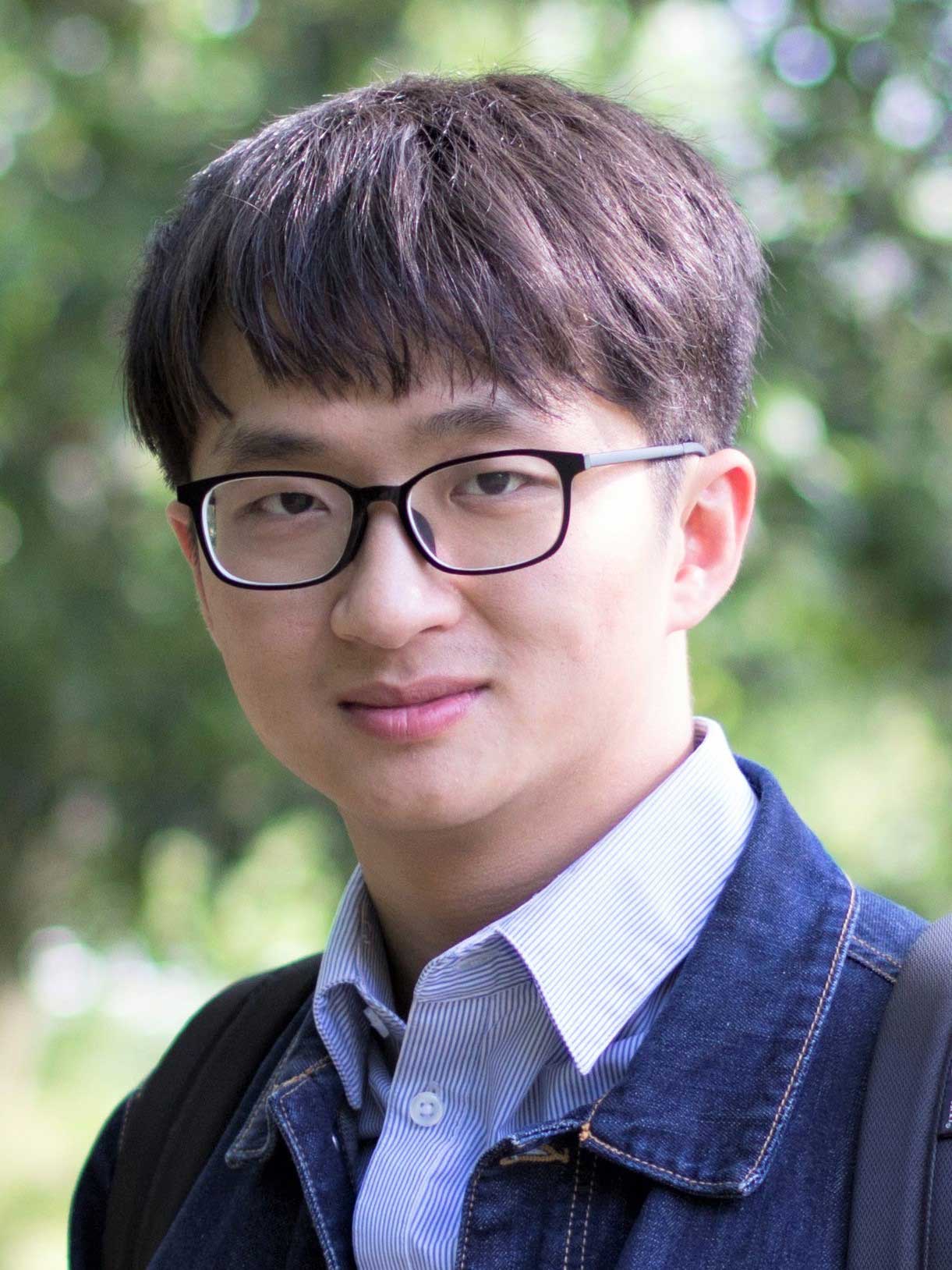}}]{Yue Zhu}
received the B.Eng and M.Sc degrees in electrical engineering, Zhejiang University, Hangzhou, China, in 2016 and 2019 respectively. He is currently a Ph.D student at Department of Electrical and Electronic Engineering, Imperial College London. His present research focuses on impedance-based stability analysis of power systems, and the noise evaluation for impedance measurement.
\end{IEEEbiography}

\vspace{-380pt}
\begin{IEEEbiography}[{\includegraphics[width=1in,height=1.25in,clip,keepaspectratio]{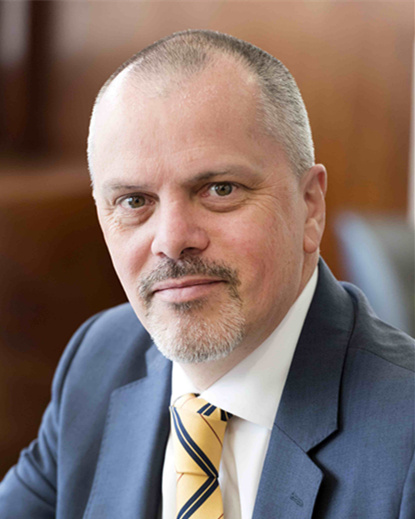}}]{Timothy C. Green}
(M'89-SM'02-F'19) received a B.Sc. (Eng) (first class honours) from Imperial College London, UK in 1986 and a Ph.D. from Heriot-Watt University, Edinburgh, UK in 1990. He is a Professor of Electrical Power Engineering at Imperial College London, and Director of the Energy Futures Lab with a role fostering interdisciplinary energy research. His research interest is in using the flexibility of power electronics to accommodate new generation patterns and new forms of load, such as EV charging, as part of the emerging smart grid. In HVDC he has contributed converter designs that reduce losses while also providing control functions assist AC system integration. In distribution systems, he has pioneered the use of soft open points and the study of stability of grid connected inverters. Prof. Green is a Chartered Engineering the UK and a Fellow of the Royal Academy of Engineering.
\end{IEEEbiography}

\end{document}